\documentclass[a4paper,11pt]{article}
\pdfoutput=1

\usepackage{jheppub} 
\usepackage{natbib}
\usepackage[utf8]{inputenc}

\usepackage{bbold}

\usepackage{graphicx}
\usepackage{amsmath,amsfonts,amssymb,amsbsy}
\usepackage{mathtools}
\usepackage{physics}

\usepackage{hyperref}
\usepackage{xcolor}
\hypersetup{
    colorlinks=true,
    pdftitle={Leading-Color Two-Loop QCD Corrections for Three-Jet Production at Hadron Colliders},
    allcolors={blue!60!black}
}

\usepackage{multirow}
\usepackage{stackrel}
\usepackage{tikz,pgfplots,pgfmath}
\usetikzlibrary{arrows.meta,decorations.markings,calc,intersections}

\usepackage{ifdraft}
\usepackage[]{easy-todo}

\usepackage{placeins}
\usepackage{slashed} 

\usepackage{booktabs}
\usepackage{adjustbox}
\usepackage{array}

\usepackage{cleveref}

\usepackage{subcaption}

\newcommand{\eps}{\epsilon}

\def\NC{{N_c}}
\def\NF{{N_f}}
\def\CA{\mathcal{A}}

\newcommand\Caravel{{\textsc{Caravel}}}

\def\Mathematica{{\sc Mathematica}}

\def\trFive{{\rm tr}_5}

\newcommand{\eqnDiag}[1]{ \vcenter{\hbox{ #1}} }

\newcommand{\ii}{\,\mathrm{i}\,}

\newcommand{\gluon}{\mathrm{g}}

\begin{document}

\title{
    Leading-Color Two-Loop QCD Corrections for Three-Jet Production at Hadron Colliders
}
\preprint{{\footnotesize \begin{tabular}{l} CERN-TH-2021-023\\ FR-PHENO-2021-06 \\ MPP-2021-20 \end{tabular}}}

\author[a,b]{S.~Abreu,}
\author[c]{F.~Febres Cordero,}
\author[d]{H.~Ita,}
\author[e]{B.~Page,}
\author[f]{and V.~Sotnikov}

\emailAdd{samuel.abreu@cern.ch}
\emailAdd{ffebres@hep.fsu.edu}
\emailAdd{harald.ita@physik.uni-freiburg.de}
\emailAdd{bpage@ipht.fr}
\emailAdd{sotnikov@mpp.mpg.de}

\affiliation[a]{Theoretical Physics Department, CERN, 1211 Geneva 23, Switzerland}
\affiliation[b]{Mani L.~Bhaumik Institute for Theoretical Physics,
Department of Physics and Astronomy, UCLA, Los Angeles, CA 90095, USA}
\affiliation[c]{Physics Department, Florida State University, 77 Chieftan Way,
Tallahassee, FL 32306, U.S.A.}
\affiliation[d]{Physikalisches Institut, Albert-Ludwigs-Universit\"at Freiburg,
Hermann-Herder-Str.~3, D-79104 Freiburg, Germany}
\affiliation[e]{Laboratoire  de  physique  de  l’Ecole  normale  sup\'erieure,
  ENS, Universit\'e  PSL,  CNRS,  Sorbonne  Universit\'e,  Universit\'e
  Paris-Diderot, Sorbonne  Paris  Cit\'e,  24  rue  Lhomond,  75005  Paris,  France}
\affiliation[f]{Max Planck Institute for Physics (Werner Heisenberg Institute),
D--80805 Munich, Germany}

\abstract{
We present the complete set of leading-color two-loop contributions required to
obtain next-to-next-to-leading-order (NNLO) QCD corrections 
to three-jet production at hadron colliders. We obtain analytic expressions
for a generating set of finite remainders, valid in the physical region
for three-jet production.
The analytic continuation of the known Euclidean-region results 
is determined from a small set of numerical evaluations of the amplitudes.
We obtain analytic expressions that are suitable for phenomenological applications and 
we present a \texttt{C++} library for their efficient and stable numerical evaluation.
}

\maketitle



\section{Introduction}\label{sec:intro}

Multi-jet events are ubiquitous at high-energy hadron
colliders such as the Large Hadron Collider (LHC). Their precise theoretical
description including higher-order quantum corrections offers a robust path to
test the Standard Model of particle physics.
This has been demonstrated, for example, by comparing measurements of the strong coupling $\alpha_s$ 
extracted from ratios between three- and two-jet production cross sections with theoretical predictions 
computed at next-to-leading order (NLO) in QCD~\cite{Abazov:2012lua,Chatrchyan:2013txa,CMS:2014mna,ATLAS:2015yaa}.
Given the amount of experimental data that has been collected since these studies were conducted and
the amount of data forecast to be collected in the near future at the LHC, there is great interest in 
obtaining theoretical predictions that will allow these types of studies to be carried out at a new
level of precision. In this paper, we provide a key missing ingredient required to reach
this goal, namely the leading-color two-loop contributions to
the three-jet production cross section at hadron colliders at
next-to-next-to-leading order (NNLO) in QCD. These are the so-called 
\textit{double-virtual} $2\rightarrow 3$ contributions, which together with the 
\textit{real-virtual} $2\rightarrow 4$ and the \textit{double-real}
$2\rightarrow 5$ contributions complete the $\mathcal{O}(\alpha_s^2)$ corrections
to the production of three jets at hadron colliders.

The calculation of NNLO QCD corrections to two-jet production at hadron
colliders has been a major success of the field in recent years. Results were obtained 
first at leading color~\cite{Ridder:2013mf,Currie:2016bfm,Currie:2017eqf} 
and later subleading-color terms were incorporated~\cite{Czakon:2019tmo} (see
also~\cite{Currie:2013dwa}). Three-jet production is a more complex process.
Results are currently known at NLO in QCD~\cite{Nagy:2001fj,Nagy:2003tz} (see
also~\cite{Kilgore:1996sq,Kilgore:2000dr}) and also including electroweak
corrections~\cite{Frederix:2018nkq,Reyer:2019obz}. Work on the NNLO QCD
corrections to three-jet production has so far been reported only for the 
leading-color two-loop scattering amplitudes, first through numerical 
evaluations~\cite{Badger:2013gxa, Badger:2017jhb, Abreu:2017hqn, 
Badger:2018gip, Abreu:2018jgq} and then in analytic
form~\cite{Gehrmann:2015bfy,Badger:2018enw,Abreu:2018zmy,Abreu:2019odu,DeLaurentis:2020qle}. 
Progress in including subleading-color contributions has been made for related 
supersymmetric theories~\cite{Abreu:2018aqd,Chicherin:2018yne,Caron-Huot:2020vlo} as well as for special 
external gluon states~\cite{Badger:2019djh}. 
In this work we present the leading-color double-virtual contributions
to the scattering cross section, starting from the known analytic 
amplitudes~\cite{Abreu:2019odu} which we analytically continue from the 
(non-physical) Euclidean region to the physical region of phase space.
Furthermore, we implement our expressions in a publicly available 
\texttt{C++} library that allows one to evaluate these contributions efficiently
and in a stable manner, making our results readily available for a future full
calculation of the corresponding NNLO QCD corrections.\footnote{The subleading-color 
contributions that we neglect are expected to have a minor
impact in phenomenological applications~\cite{Berger:2009ep,Ita:2011ar,Currie:2013dwa,Czakon:2019tmo}.}

Our work builds on a number of recent technical advances that have paved the way to the calculation of two-loop scattering amplitudes for $2\rightarrow 3$
processes with phenomenological
applications~\cite{Abreu:2020cwb,Chawdhry:2020for,Chawdhry:2019bji,
Kallweit:2020gcp,Agarwal:2021grm,Badger:2021nhg}.
We employ the known pure two-loop master
integrals~\cite{Papadopoulos:2015jft,Gehrmann:2015bfy,Abreu:2018aqd,Chicherin:2018old}, which
can be written in  terms of a special set of multi-valued transcendental
functions---the so-called pentagon functions~\cite{Gehrmann:2018yef,Chicherin:2020oor}.  
In ref.~\cite{Chicherin:2020oor}, the five-point integrals are expressed
in terms of a basis of functions for all assignments of the incoming
momenta required for the double-virtual corrections we compute, and they
can be efficiently evaluated in the physical region corresponding to three-jet production.

We obtain the analytic form of a generating set of partial amplitudes in the physical region
by analytically continuing the known amplitudes from the Euclidean region~\cite{Abreu:2019odu}.
Based on considerations on the analytic continuation of the pentagon functions,
the continuation of each helicity amplitude is reduced to a linear-algebra problem
that can be efficiently solved using
evaluations of the amplitudes in a finite field~\cite{vonManteuffel:2014ixa,Peraro:2016wsq,Abreu:2017hqn}.
These evaluations are performed with the program \Caravel{}~\cite{Abreu:2020xvt}. 
This is a new generic procedure that allows to upgrade results for amplitudes in the Euclidean 
region to results valid in other regions.
Finally, we assemble all partial amplitudes into squared color- and
helicity-summed finite remainders, which can be numerically evaluated over
phase space with the provided \texttt{C++} library~\cite{FivePointAmplitudes}. 
We perform several checks on our results and verify 
the efficiency and numerical stability of our implementation.

The paper is organized as follows. In \cref{sec:notAndConv} we establish our
conventions and define the objects we will be computing. In \cref{sec:method}
we describe our procedure to compute the partial amplitudes
in the physical region and the different permutations that are required
to obtain squared finite remainders.
In \cref{sec:results} we present our results in the form of ancillary files and 
showcase the efficiency and numerical stability of our
code for the numerical evaluation of the double-virtual NNLO contributions
to the production of three jets at hadron colliders, and
discuss the checks that we have performed on our results. We summarize our results
in~\cref{sec:conclusion}.


\section{Notation and Conventions}\label{sec:notAndConv}

\subsection{Helicity Amplitudes}

We consider all the channels contributing to the production of three partons
at hadron colliders. We label the initial-state particles
with indices 1 and 2, and the final-state particles with indices 3, 4 and 5.
There is a single channel with five gluons,
\begin{equation}\label{eq:gProc}
  \gluon^{h_1}_{p_1}\gluon^{h_2}_{p_2}\to \gluon^{h_3}_{p_3}\gluon^{h_4}_{p_4}\gluon^{h_5}_{p_5}\,,
\end{equation}
three channels with a pair of quark and anti-quark,
\begin{equation}\label{eq:qProc}
  \bar{q}^{h_1}_{p_1}q^{h_2}_{p_2}\to \gluon^{h_3}_{p_3}\gluon^{h_4}_{p_4}\gluon^{h_5}_{p_5}\,,\quad 
  \bar{q}^{h_1}_{p_1}\gluon^{h_2}_{p_2}\to\bar{q}^{h_3}_{p_3}\gluon^{h_4}_{p_4}\gluon^{h_5}_{p_5}\,,\quad 
  \gluon^{h_1}_{p_1}\gluon^{h_2}_{p_2}\to q^{h_3}_{p_3}\bar{q}^{h_4}_{p_4}\gluon^{h_5}_{p_5}\,,
\end{equation}
and four channels with two distinct pairs of quark and anti-quark
\begin{align}\begin{split}\label{eq:qQProc}
  &\bar{q}^{h_1}_{p_1}q^{h_2}_{p_2}\to Q^{h_3}_{p_3}\bar{Q}^{h_4}_{p_4}\gluon^{h_5}_{p_5}\,,\quad 
  \bar{q}^{h_1}_{p_1}Q^{h_2}_{p_2}\to Q^{h_3}_{p_3}\bar{q}^{h_4}_{p_4}\gluon^{h_5}_{p_5}\,,\\
  &\bar{q}^{h_1}_{p_1}\bar{Q}^{h_2}_{p_2}\to\bar{q}^{h_3}_{p_3}\bar{Q}^{h_4}_{p_4}\gluon^{h_5}_{p_5}\,,\quad 
  \bar{q}^{h_1}_{p_1}\gluon^{h_2}_{p_2}\to\bar{q}^{h_3}_{p_3}Q^{h_4}_{p_4}\bar{Q}^{h_5}_{p_5}.
\end{split}\end{align}
In \cref{eq:gProc,eq:qProc,eq:qQProc}, we identify each particle by its type ($\gluon$ for gluon, 
$q$ or $Q$ for distinct quarks, and $\bar{q}$ and $\bar{Q}$ for the corresponding anti-quarks),
a superscript $h_i$ for the helicity state and a subscript $p_i$ for the four momentum.
All other channels involving quarks are related to the ones above by charge conjugation
and permutation of incoming or outgoing labels. While in \cref{eq:qQProc}, we assumed the two
quark pairs to be of distinct flavours, we should also consider the case with two
quark lines of identical flavour. There are three such channels which, schematically, can be obtained from
\begin{align}\label{eq:4qProc}\begin{split}
  \bar{q}^{h_1}_{p_1}q^{h_2}_{p_2}\to q^{h_3}_{p_3}\bar{q}^{h_4}_{p_4}\gluon^{h_5}_{p_5}=
  \left(\bar{q}^{h_1}_{p_1}q^{h_2}_{p_2}\to Q^{h_3}_{p_3}\bar{Q}^{h_4}_{p_4}\gluon^{h_5}_{p_5}\right)-
  \left(\bar{q}^{h_1}_{p_1}Q^{h_2}_{p_2}\to Q^{h_3}_{p_3}\bar{q}^{h_4}_{p_4}\gluon^{h_5}_{p_5}\right)\,,\\
  \bar{q}^{h_1}_{p_1}\bar{q}^{h_2}_{p_2}\to\bar{q}^{h_3}_{p_3}\bar{q}^{h_4}_{p_4}\gluon^{h_5}_{p_5}=
  \left(\bar{q}^{h_1}_{p_1}\bar{Q}^{h_2}_{p_2}\to\bar{q}^{h_3}_{p_3}\bar{Q}^{h_4}_{p_4}\gluon^{h_5}_{p_5}\right)-
  \left(\bar{q}^{h_1}_{p_1}\bar{Q}^{h_2}_{p_2}\to\bar{Q}^{h_3}_{p_3}\bar{q}^{h_4}_{p_4}\gluon^{h_5}_{p_5}\right)\,,\\
  \bar{q}^{h_1}_{p_1}\gluon^{h_2}_{p_2}\to\bar{q}^{h_3}_{p_3}q^{h_4}_{p_4}\bar{q}^{h_5}_{p_5}=
  \left(\bar{q}^{h_1}_{p_1}\gluon^{h_2}_{p_2}\to\bar{q}^{h_3}_{p_3}Q^{h_4}_{p_4}\bar{Q}^{h_5}_{p_5}\right)-
  \left(\bar{q}^{h_1}_{p_1}\gluon^{h_2}_{p_2}\to\bar{Q}^{h_3}_{p_3}Q^{h_4}_{p_4}\bar{q}^{h_5}_{p_5}\right)\,.
\end{split}\end{align}
In order to obtain them we thus also compute the (redundant) channels
\begin{equation}\label{eq:qQProcP}
  \bar{q}^{h_1}_{p_1}\bar{Q}^{h_2}_{p_2}\to\bar{Q}^{h_3}_{p_3}\bar{q}^{h_4}_{p_4}\gluon^{h_5}_{p_5}\,,\quad 
  \bar{q}^{h_1}_{p_1}\gluon^{h_2}_{p_2}\to\bar{Q}^{h_3}_{p_3}Q^{h_4}_{p_4}\bar{q}^{h_5}_{p_5}.
\end{equation}
We remark that, when labelling channels, such as in
\cref{eq:qQProcP,eq:4qProc,eq:qQProc,eq:qProc,eq:gProc}, the momenta $p_1$ and $p_2$ are
taken to be incoming and $p_3, p_4$ and $p_5$ are taken to be outgoing. In all
other contexts we use the symmetric ``all-outgoing'' convention of
refs.~\cite{Abreu:2018jgq,Abreu:2019odu}, such that momentum conservation
is $p_1+p_2+p_3+p_4+p_5=0$.

We will consider each of the channels above in the leading-color approximation.
Let us denote the set of helicities by a vector $\vb*{h}$ and
the set of momenta by a vector $\vb*{p}$. The amplitudes $\mathcal{M}(\vb*{h},\vb*{p})$
describing each of the channels above,
which are vectors in color space,
can be expanded in powers of the bare strong 
coupling $\alpha_s^0=(g_s^0)^2/(4\pi)$,
\begin{equation}\label{eq:pertExp}
  \mathcal{M} = \left(g_s^0\right)^3\left(\mathcal{M}^{(0)}
  +\frac{\alpha^{0}_{s}}{2\pi}\mathcal{M}^{(1)}
  +\left(\frac{\alpha^{0}_{s}}{2\pi}\right)^2\mathcal{M}^{(2)}
  +\mathcal{O}\left((\alpha^{0}_s)^3\right)\right)\,,
\end{equation}
where we drop the dependence on $\vb*{h}$ and $\vb*{p}$ for compactness. 
The order of $\alpha^0_s$ is aligned with the number of loops of the Feynman diagrams contributing
to the associated coefficient, so we refer to $\mathcal{M}^{(1)}$ and $\mathcal{M}^{(2)}$
as the one-loop and two-loop amplitudes respectively, and to $\mathcal{M}^{(0)}$ as the tree-level
amplitude. 
All calculations will be done in the 't Hooft-Veltman (HV) scheme of dimensional regularization 
with $D=4-2\epsilon$ space-time dimensions and we mostly keep the dependence on $\epsilon$ implicit.
For our definition of helicity amplitudes with external quarks, we refer the reader
to the detailed discussion in ref.~\cite{Abreu:2018jgq}.

For each channel, the coefficients $\mathcal{M}^{(i)}$ can be decomposed in terms of different color structures,
and at leading color this decomposition is independent of the loop order. We write
\begin{equation}\label{eq:colourDec}
  \mathcal{M}^{(i)}(\vb*{h},\vb*{p})=\left(\frac{N_c}{2}\right)^{i}\sum_{\sigma\in\Sigma}\mathcal{C}\left(\sigma\right)
  \Phi\left(\sigma(\vb*{h}),\sigma(\vb*{p})\right)
  \mathcal{A}^{(i)}\left(\sigma(\vb*{h}),\sigma(\vb*{p})\right)\,,
\end{equation}
where $N_c$ is the number of colors, $\mathcal{C}\left(\sigma\right)$ denotes a vector in color space and
the sum is over a channel-dependent set of permutations $\Sigma$.
We defer a more detailed discussion of the sets $\Sigma$ and the vectors 
$\mathcal{C}\left(\sigma\right)$ to \cref{sec:partialsAssemble}. For now,
we simply define the action of a permutation 
$\sigma = \{\sigma_1,\sigma_2,\sigma_3,\sigma_4,\sigma_5\} \in S_5$ on 
momenta and helicity labels as $\sigma(p_i) \coloneqq  p_{\sigma_i}$ and 
$\sigma(h_i) \coloneqq  h_{\sigma_i}$. The $\Phi\left(\sigma(\vb*{h}),\sigma(\vb*{p})\right)$
are normalization factors, and the $\mathcal{A}^{(i)}\left(\sigma(\vb*{h}),\sigma(\vb*{p})\right)$
are partial amplitudes.

For each term in the sum, 
the coefficient of each $\mathcal{C}(\sigma)$ has a fixed ordering of the external momenta 
determined by the permutation $\sigma$.
The  $\Phi\left(\sigma(\vb*{h}),\sigma(\vb*{p})\right)$ are defined  so that
$\mathcal{A}^{(0)}\left(\sigma(\vb*{h}),\sigma(\vb*{p})\right)\equiv1$ whenever the tree-level
amplitude is non-vanishing, and for the other cases we choose\footnote{Here we use the same notation for spinor brackets as in \cite{Abreu:2020xvt}.}
\begin{subequations}\label{eq:spinor-weights}
  \begin{align}
    \Phi(\vb*{h}=\{+1,+1,+1,+1,+1\},\,\vb*{p}) & \coloneqq  \frac{[2 1]^2 \langle 1 2\rangle ^3 \langle 1 3\rangle }
    {\langle 1 4\rangle ^2 \langle 1 5\rangle ^2 \langle 2 3\rangle ^3}, \\
    \Phi(\vb*{h}=\{-1,+1,+1,+1,+1\},\,\vb*{p}) & \coloneqq  \frac{[2 1] \langle 1 2\rangle ^4 \langle 1 3\rangle ^3}
    {\langle 1 4\rangle ^2 \langle 1 5\rangle ^2 \langle 2 3\rangle ^5}, \\
    \Phi(\vb*{h}=\{+\frac{1}{2},-\frac{1}{2},+1,+1,+1\},\,\vb*{p}) & \coloneqq  
    \frac{[3 1] \langle 1 2\rangle ^3 \langle 1 3\rangle }
    {\langle 1 4\rangle ^2 \langle 1 5\rangle ^2 \langle 2 3\rangle ^2},
  \end{align}
\end{subequations}
where we set $\vb*{p}=\{p_1,p_2,p_3,p_4,p_5\}$. With this choice,
the partial amplitudes $\mathcal{A}\left(\sigma(\vb*{h}),\sigma(\vb*{p})\right)$
only depend on the kinematics through the Mandelstam invariants
\begin{align}\begin{split}\label{eq:mandDef}
  s_{12}=(p_1+&p_2)^2\,,\quad
  s_{23}=(p_2+p_3)^2\,,\quad
  s_{34}=(p_3+p_4)^2\,,\quad\\
  &s_{45}=(p_4+p_5)^2\,,\quad
  s_{15}=(p_1+p_5)^2\,,\quad
\end{split}\end{align}
which we collectively denote as $\vec s=\{s_{12},s_{23},s_{34},s_{45},s_{15}\}$,
and the parity-odd contraction of four momenta
\begin{equation}\label{eq:tr5}
  \trFive \coloneqq 4\ii\varepsilon(p_1,p_2,p_3,p_4),
\end{equation}
where $\varepsilon(\boldsymbol\cdot,\boldsymbol\cdot,\boldsymbol\cdot,\boldsymbol\cdot )$ is the fully anti-symmetric Levi-Civita tensor, 
whose sign is fixed such that $\Im(\trFive)>0$ corresponds to the choice of the basis helicity amplitudes in ref.~\cite{Abreu:2019odu}.
The action of $\sigma$ on quantities that depend on momenta is inherited from
the action on momentum labels. For instance, the action on
the Mandelstam invariants $s_{ij}=(p_i+p_j)^2$ is given by
$\sigma(s_{ij}) = s_{\sigma_i\sigma_j}$, and on the parity-odd invariant in \cref{eq:tr5} by
$\sigma(\trFive) = \mathrm{sgn}(\sigma)\,\trFive$.

\begin{figure}
  \centering
  \begin{align*}
        &\mathcal{A}^{(2)}_{\gluon\gluon \rightarrow \gluon\gluon\gluon} \sim
          \eqnDiag{\includegraphics[scale=0.38]{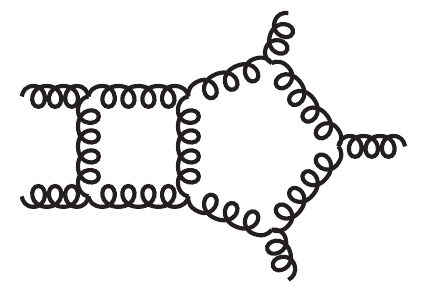}} +
          \eqnDiag{\includegraphics[scale=0.38]{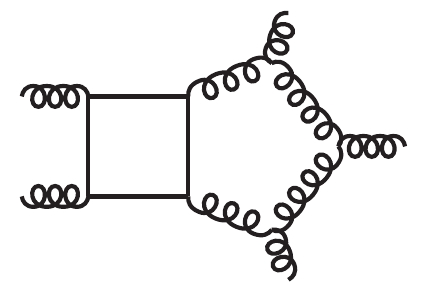}} +
          \eqnDiag{\includegraphics[scale=0.38]{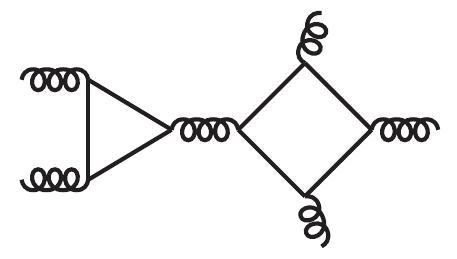}} + \ldots, \\
        &\mathcal{A}^{(2)}_{\bar{q}q \rightarrow \gluon\gluon\gluon} \sim
          \eqnDiag{\includegraphics[scale=0.38]{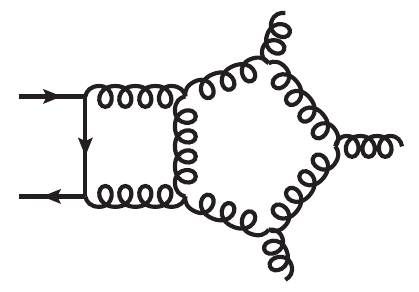}} +
          \eqnDiag{\includegraphics[scale=0.38]{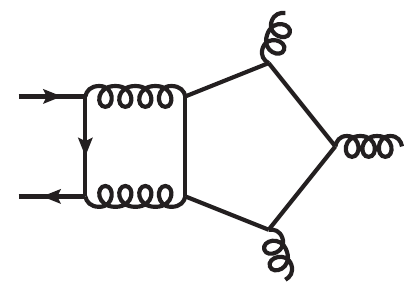}} +
          \eqnDiag{\includegraphics[scale=0.38]{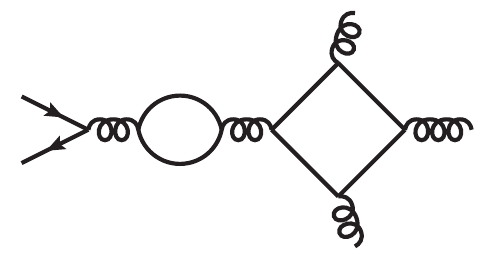}} + \ldots, \\
        &\mathcal{A}^{(2)}_{\bar{q}q \rightarrow Q\bar{Q}\gluon} \sim
          \eqnDiag{\includegraphics[scale=0.38]{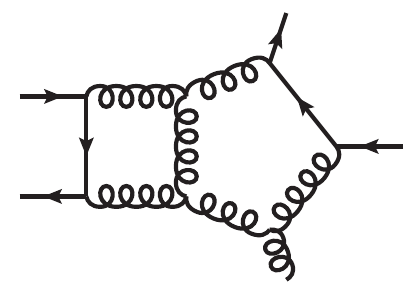}} +
          \eqnDiag{\raisebox{5mm}{\includegraphics[scale=0.38]{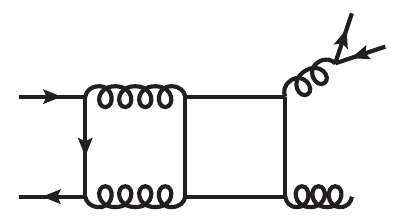}}} +
          \eqnDiag{\raisebox{4mm}{\includegraphics[scale=0.38]{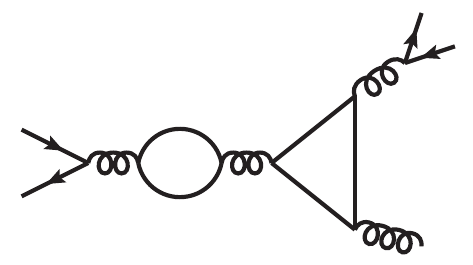}}}  + \ldots.
  \end{align*}
  
  \caption{Characteristic Feynman diagrams which contribute to 
    $\mathcal{A}^{(2)[j]}$ for representative channels with no external quark lines, a single
    quark line and two external quark lines, for $j=0$~(left column), $j=1$ (middle column) and $j=2$ (right column).}
  \label{fig:ampCharacteristic}
\end{figure}

We work in the leading-color approximation where the number of colors $N_c$ is large
and the ratio $\NF/\NC$ is kept fixed, where $\NF$ denotes the number of massless quark flavors.
In this limit, each of the $\mathcal{A}^{(i)}\left(\vb*{h},\vb*{p}\right)$ can be further decomposed
in powers of $\NF$. We write
\begin{equation}\label{eq:nfDec}
  \mathcal{A}^{(i)}\left(\vb*{h},\vb*{p}\right)=
  \sum_{j=0}^{i}\left(\frac{N_f}{N_c}\right)^j
  \mathcal{A}^{(i)[j]}\left(\vb*{h},\vb*{p}\right)\,.
\end{equation}
In \cref{fig:ampCharacteristic} we depict 
characteristic diagrams that contribute to $\mathcal{A}^{(2)[j]}$ for
representative channels with no external quark lines, 
a single external quark line or two external quark lines.

The renormalized amplitudes $\mathcal{M}_R$ are obtained from the bare amplitudes $\mathcal{M}$ 
by considering an expansion in powers of the renormalized coupling $\alpha_s(\mu)$.
In the $\overline{\text{MS}}$-scheme, the later is related to the bare coupling $\alpha_s^0$ through
\begin{equation}\label{eq:renormCoupling}
    \alpha_0\mu_0^{2\epsilon}S_{\epsilon}
  =\alpha_s(\mu)\mu^{2\epsilon}\left(
  1-\frac{\beta_0}{\epsilon}\frac{\alpha_s(\mu)}{2\pi}
  +\left(\frac{\beta_0^2}{\epsilon^2}-\frac{\beta_1}{2\epsilon}\right)
  \left(\frac{\alpha_s(\mu)}{2\pi}\right)^2+\mathcal{O}
  \left(\alpha_s^3(\mu)\right)\right),
\end{equation}
where $S_\epsilon=(4\pi)^{\epsilon}e^{-\epsilon\gamma_E}\,$, and $\mu_0$ and $\mu$ are the dimensional
regularization and renormalization scales respectively, which from now on we assume to be equal. 
We will also suppress the $\mu$ dependence in the coupling and write 
$\alpha_s\equiv\alpha_s(\mu)$.
The $\beta_i$ are the coefficients of the QCD $\beta$-function,
\begin{equation}\label{eq:betai}
  \begin{aligned}
    \beta_0&=\frac{11 C_A - 4 T_F \NF }{6}=\frac{11N_c-2\NF}{6} \,,\quad\\
    \beta_1&=\frac{17C_A^2-6C_FT_F\NF-10C_AT_F\NF}{6}
    =\frac{1}{6}\left(17N_c^2-\frac{13}{2}N_c\NF\right)+\mathcal{O}(N_c^{-1})\,,
  \end{aligned}
\end{equation}
where we have set $T_F=1/2$, $C_A=N_c$ and $C_F=(N_c^2-1)/(2N_c)$, and only kept
terms that contribute at leading color.
The renormalized partial amplitudes are related to their bare counterparts as
\begin{align}\begin{split}
  \label{eq:twoLoopUnRenorm}
    &\mathcal{A}_R^{(0)}=\mathcal{A}^{(0)}, \\
  & \mathcal{A}_R^{(1)}=S_{\epsilon}^{-1}\mathcal{A}^{(1)}
  -\frac{3\beta_0}{\epsilon N_c}
  \mathcal{A}^{(0)}\,,\\
  &\mathcal{A}_R^{(2)}=
  S_{\epsilon}^{-2}\mathcal{A}^{(2)}
  -\frac{5\beta_0}{\epsilon N_c}
  S_{\epsilon}^{-1}
  \mathcal{A}^{(1)}
  +\left(\frac{15\beta_0^2}{2\epsilon^2N_c^2}
  -\frac{3\beta_1}{\epsilon N_c^2}\right)
  \mathcal{A}^{(0)}\,.
\end{split}\end{align}

\subsection{Finite Remainders}

Amplitudes contribute to physical observables only through so-called finite 
remainders~(see e.g.\ \cite{Weinzierl:2011uz}),
which we denote $\mathcal{R}\left(\vb*{h},\vb*{p}\right)$.
We define a remainder for each partial amplitude $\mathcal{A}\left(\vb*{h},\vb*{p}\right)$.
They are obtained from the renormalized amplitudes by removing the remaining singularities
of infrared origin that can be determined from lower-loop amplitudes and well-known
universal factors~\cite{Catani:1998bh,Becher:2009cu,Gardi:2009qi}.
The finite remainders can be expanded in powers
of~$\alpha_s$,
\begin{equation}
  \mathcal{R} = \mathcal{R}^{(0)} + \frac{\alpha_s}{2\pi} \mathcal{R}^{(1)} + \left(\frac{\alpha_s}{2\pi}\right)^2 \mathcal{R}^{(2)} + \mathcal{O}(\alpha_s^3)\,,
\end{equation}
with the $\mathcal{R}^{(i)}$ defined as
\begin{align}\begin{split} \label{eq:remainder2l}
    \mathcal{R}^{(0)} &= \mathcal{A}_{R}^{(0)}, \\
    \mathcal{R}^{(1)} &= \mathcal{A}_R^{(1)}-\mathbf{I}^{(1)}\mathcal{A}_R^{(0)} ~+\mathcal{O}(\epsilon), \\ 
    \mathcal{R}^{(2)} &= \mathcal{A}_R^{(2)}-\mathbf{I}^{(1)}\mathcal{A}_R^{(1)}
    -\mathbf{I}^{(2)}{\cal A}_R^{(0)} ~+\mathcal{O}(\epsilon),
\end{split}\end{align}
The operators ${\bf I}^{(i)}$ are channel specific, and we collect them in
\cref{sec:appIR} for all the channels in \cref{eq:gProc,eq:qProc,eq:qQProc}.
In these expressions, we extend the expansion of 
${\bf I}^{(1)}\mathcal{A}_R^{(1)}$ and
${\bf I}^{(2)}\mathcal{A}_R^{(0)}$ to also include
terms of order $\eps^0$. This subtracts non-trivial contributions from
the finite term of $\mathcal{A}_R^{(1)}$ and $\mathcal{A}_R^{(2)}$ that are related to 
the lower-loop amplitudes. It is clear that the finite remainders $\mathcal{R}^{(k)}$ contain
the genuine new information at order $k$. 
We stress that even though our calculations are performed in the HV scheme of dimensional regularization, 
the remainders computed in this scheme agree with the ones computed in the 
conventional dimensional regularization (CDR) scheme, see e.g.~\cite{Broggio:2015dga}.
Finally, the finite remainders can also be 
decomposed into powers of $\NF$, similarly to \cref{eq:nfDec},
\begin{equation} \label{eq:remainder-nf-contrib}
\mathcal{R}^{(i)}\left(\vb*{h},\vb*{p}\right)=
  \sum_{j=0}^{i}\left(\frac{N_f}{N_c}\right)^j
  \mathcal{R}^{(i)[j]}\left(\vb*{h},\vb*{p}\right)\,.
\end{equation}
Since the functions $\mathcal{R}^{(i)[j]}$ are dimensionless by construction, their scale dependence
can be restored at any point by rescaling the momenta as $\vb*{p}\to \vb*{p}/\mu$ (we set the factorization scale $\mu_F$
to be equal to the renormalization scale, $\mu_F=\mu$).
Unless it happens to be important for the point we wish to make, we will keep this dependence implicit.

The color-dressed remainder $\mathcal{R}^{(i)}_\mathcal{M}\left(\vb*{h},\vb*{p}\right)$ is 
trivially obtained from the decomposition of the color-dressed amplitude $\mathcal{M}^{(i)}(\vb*{h},\vb*{p})$ into partial 
amplitudes, see \cref{eq:colourDec}:
\begin{align}\begin{split}\label{eq:colourDecRem}
  \mathcal{R}_\mathcal{M}(\vb*{h},\vb*{p})&=g_s^3\left(\mathcal{R}_\mathcal{M}^{(0)}(\vb*{h},\vb*{p})
  +\frac{\alpha_s}{2\pi}\mathcal{R}_\mathcal{M}^{(1)}(\vb*{h},\vb*{p})+\left(\frac{\alpha_s}{2\pi}\right)^2
  \mathcal{R}_\mathcal{M}^{(2)}(\vb*{h},\vb*{p})+\mathcal{O}(\alpha_s^3)\right)\,,\\
  \mathcal{R}_\mathcal{M}^{(i)}(\vb*{h},\vb*{p})&=\left(\frac{N_c}{2}\right)^{i}\sum_{\sigma\in\Sigma}\mathcal{C}\left(\sigma\right)
  \Phi\left(\sigma(\vb*{h}),\sigma(\vb*{p})\right)
  \mathcal{R}^{(i)}\left(\sigma(\vb*{h}),\sigma(\vb*{p})\right)\,.
\end{split}\end{align}
Whenever the distinction is important, we will call the $\mathcal{R}^{(i)}\left(\sigma(\vb*{h}),\sigma(\vb*{p})\right)$
partial remainders, in analogy with the concept of partial amplitudes. We also recall that, even though this is kept implicit,
the color-dressed remainders are vectors in color space (in general, the $\mathcal{C}\left(\sigma\right)$ have open (anti-)fundamental 
and adjoint color indices, see \cref{sec:partialsAssemble} for their explicit form).

\subsection{Squared Finite Remainders}\label{sec:sqR}
\begin{figure}
  \centering
  \begin{subfigure}{0.45\textwidth}\centering
    \includegraphics[scale=0.5]{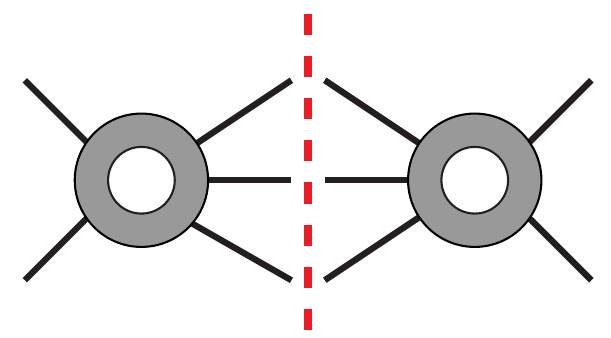}
  \end{subfigure}
  \begin{subfigure}{0.45\textwidth}\centering
    \includegraphics[scale=0.5]{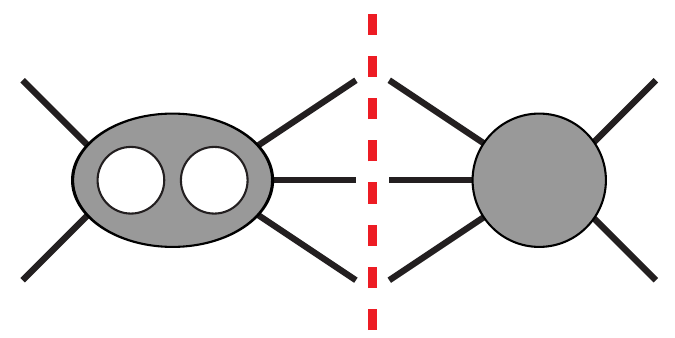}
  \end{subfigure}
  \caption{Schematic representation for the two kinds of contributions to $H^{(2)}$: the square of one-loop amplitudes,
  and the interference between tree-level and two-loop amplitudes. }
  \label{fig:h2contributions}
\end{figure}

 When computing physical observables, we are interested in the squared remainder,
summed over color states and, in the case of unpolarized observables, over helicity states. 
We denote this object as $H$. It admits an expansion in powers of $\alpha_s$ as
\begin{equation}\label{e:hDef}
  H=H^{(0)}+\frac{\alpha_s}{2\pi}H^{(1)}+\left(\frac{\alpha_s}{2\pi}\right)^2H^{(2)}+\mathcal{O}(\alpha_s^3).
\end{equation}
In terms of the quantities in \cref{eq:colourDecRem}, $H$ is defined as
\begin{equation}
  H= \frac{1}{\mathcal{N}_H}\,\sum_{\vb*{h}}\mathcal{R}^\dagger_{\mathcal{M}}(\vb*{h},\vb*{p})\mathcal{R}_{\mathcal{M}}(\vb*{h},\vb*{p})\,,
\end{equation}
where the normalization $\mathcal{N}_H$ is fixed such that $H^{(0)}=1$, 
\begin{equation}
  \mathcal{N}_H \coloneqq (4\pi\alpha_s)^3\sum_{\vb*{h}} \abs{\mathcal{R}^{(0)}_\mathcal{M}(\vb*{h},\vb*{p})}^2\,.
\end{equation}
To keep the notation light, in this equation and the following 
we do not write explicitly the sum over contracted colour indices which is understood to be performed
in the squaring procedure.
At order $\alpha_s$ we have
\begin{equation}\label{eq:h1}
  H^{(1)}=\frac{1}{\sum_{\vb*{h}} \abs{\mathcal{R}^{(0)}_\mathcal{M}(\vb*{h},\vb*{p})}^2}
  \sum_{\vb*{h}} \, 
  2\Re\left[\mathcal{R}^{(0)\dagger}_\mathcal{M}(\vb*{h},\vb*{p})\mathcal{R}^{(1)}_\mathcal{M}(\vb*{h},\vb*{p})\right]\,,
\end{equation} 
and at order $\alpha^2_s$
\begin{align}\label{eq:h2}
  H^{(2)}=\frac{1}{\sum_{\vb*{h}} \abs{\mathcal{R}^{(0)}_\mathcal{M}(\vb*{h},\vb*{p})}^2}
  \left(\sum_{\vb*{h}} \abs{\mathcal{R}^{(1)}_\mathcal{M}(\vb*{h},\vb*{p})}^2
  +\sum_{\vb*{h}}  2\Re\left[\mathcal{R}^{(0)\dagger}_\mathcal{M}(\vb*{h},\vb*{p})\mathcal{R}^{(2)}_\mathcal{M}(\vb*{h},\vb*{p})\right]
  \right)\,,
\end{align}
with each of the two contributions depicted in \cref{fig:h2contributions}.
We note that, when summing over contracted color indices, we have
$\mathcal{C}^{\dagger}(\sigma^\prime)~\mathcal{C}(\sigma)\sim \delta_{\sigma\sigma'}$ in the leading-color approximation
we adopt. This greatly simplifies the construction of the squared remainders.

To be clear about which terms we keep in the leading-color approximation, 
we explicitly write the decomposition of our remainders in powers of $\NF/\NC$.
For $H^{(1)}$, we have
\begin{equation}\label{eq:h1NfDec}
  H^{(1)}=\frac{\NC}{2}\left(H^{(1)[0]}+\frac{\NF}{\NC}H^{(1)[1]}\right)\,.
\end{equation}
For $H^{(2)}$,
\begin{equation}\label{eq:h2NfDec}
  H^{(2)}=\left(\frac{\NC}{2}\right)^2\left(H^{(2)[0]}+\frac{\NF}{\NC}H^{(2)[1]}+\left(\frac{\NF}{\NC}\right)^2H^{(2)[2]}
  \right)\,.
\end{equation}
For most channels the subleading-color corrections are suppressed by a factor of $1/N_c^{2}$. However, for the channels
in \cref{eq:4qProc} with two identical quark lines the suppression is by a single power of $N_c$.


\section{Physical Region Helicity Amplitudes}\label{sec:method}

Our goal in this paper is to obtain the double-virtual contributions
for three-jet production at hadron colliders at NNLO.
The central object we need to compute is the squared finite remainder
defined in \cref{sec:sqR}. In this section, we discuss how 
we construct the different contributions  starting from the Euclidean-region partial amplitudes obtained
in ref.~\cite{Abreu:2019odu}. We recall the comment below \cref{eq:qQProcP} regarding
the change of conventions for denoting each channel between this paper and ref.~\cite{Abreu:2019odu}.

\subsection{Partial Amplitudes in the Physical Region}\label{sec:physReg}

The first step towards the construction of the squared finite remainder in the physical region is to
obtain the amplitudes computed in ref.~\cite{Abreu:2019odu} in such region. 
Here we consider only the basis set of partial helicity amplitudes of
ref.~\cite{Abreu:2019odu}, and we will return to their permutations required to assemble \cref{eq:h2} in \cref{sec:partialsAssemble}.
The amplitudes of ref.~\cite{Abreu:2019odu} were computed in the
(non-physical) Euclidean region where 
\begin{equation}\label{eq:eucregion}
  s_{12},s_{23},s_{34},s_{45},s_{15} < 0\,, \qquad \trFive^2 >0\,.
\end{equation}
It is our aim to construct results in the physical region associated with the channels in
\cref{eq:gProc,eq:qProc,eq:qQProc,eq:qQProcP,eq:4qProc},
where $p_1$ and $p_2$ are initial-state momenta and $p_3$, $p_4$ and $p_5$
are final-state momenta. This region is characterized by
\begin{equation}\label{eq:physregion}
  s_{12},s_{34},s_{45} > 0, \qquad s_{23},s_{15} < 0, \qquad \trFive^2 < 0.
\end{equation}
The latter condition is trivially satisfied by real-valued momenta.
In this section, for clarity, we will denote a phase-space point in the physical region as
$\vb*{p}_\varphi$ and a phase-space point in the Euclidean region as $\vb*{p}_E$.
For region-agnostic statements we simply denote a phase-space point as
$\vb*{p}$.

In order to explain how we convert the results of ref.~\cite{Abreu:2019odu} to the kinematic region
in \cref{eq:physregion}, let us briefly summarize how they were obtained.
The discussion below holds for each power of $\NF$, that is for each
$\mathcal{A}^{(i)[j]}$, but for simplicity of the expressions we will
suppress this dependence in intermediate steps.
The first step is a computation of the decomposition of the bare partial amplitudes
at one and two loops into a set of master integrals:
\begin{equation}\label{eq:ampAsMaster}
  \mathcal{A}^{(i)}(\vb*{h},\vb*{p};\eps)=
  \sum_{k}c^{(i)}_k(\vb*{h},\vb*{p};\eps)\,m^{(i)}_k(\vb*{p};\eps)\,,\qquad i=1,2\,,
\end{equation}
where we make the dependence on the dimensional regulator $\eps$ explicit.
This decomposition was obtained using the two-loop numerical unitarity 
approach \cite{Ita:2015tya,Abreu:2017xsl,Abreu:2017hqn,Abreu:2018jgq}.
At this stage, the master integrals were replaced by their expansion in terms of 
Euclidean-region pentagon functions.\footnote{These functions are
a basis for the transcendental
functions appearing in five-point massless one- and two-loop master integrals.
We refer the reader to ref.~\cite{Gehrmann:2018yef} for more
details on their definition.}
Denoting the vector space spanned by relevant pentagon-function monomials in this region as $\vec f(\vb*{p}_E)=\{f_l(\vb*{p}_E)\}_{l\in B}$,
with $B$ denoting the set of multi-indices $l$, we write
\begin{equation}\label{eq:mastersToPent}
  m^{(i)}_k(\vb*{p}_E;\eps)=\sum_{k=0}^4\sum_{l\in B}\epsilon^k d^{(i)}_{k,l} f_l(\vb*{p}_E) + \mathcal{O}(\epsilon^5)\,.
\end{equation}
In this decomposition, we assume that we have a pure basis of master integrals \cite{ArkaniHamed:2010gh} at one 
and two loops, in which case the coefficients $d^{(i)}_{k,l}$ are rational numbers.
Furthermore, we assume all integrals are normalized so that their Laurent expansion around
$\epsilon=0$ has no poles, and stress that the set of pentagon functions is common
to one and two loops.
By inserting the decomposition \eqref{eq:mastersToPent} into \cref{eq:ampAsMaster}
for both one- and two-loop amplitudes, and then using these expressions in
the definitions for the finite remainders given in \cref{eq:remainder2l},
we obtain a decomposition of the two-loop remainders into pentagon functions:
\begin{equation}\label{eq:remAsPent}
  \mathcal{R}^{(i)[j]}(\vb*{h},\vb*{p}_E)=
  \sum_{k\in B}r^{(i)[j]}_k(\vb*{h},\vb*{p}_E)\,f_k(\vb*{p}_E)\,,
\end{equation}
where we have reintroduced the superscript $[j]$ to stress that the decomposition holds
for each power of $\NF$. Obtaining the analytic form of the coefficients
$r^{(i)[j]}_k(\vb*{h},\vb*{p})$ was the main result of ref.~\cite{Abreu:2019odu}.
Let us finish this summary by commenting on the structure of the coefficients
$r^{(i)[j]}_k(\vb*{h},\vb*{p})$. They can be written as
\begin{equation}
  r_k(\vb*{h},\vb*{p})=r^+_k(\vb*{h}, \vec s)+ r^-_k(\vb*{h}, \vec s)\,\trFive\,,
\end{equation}
where we briefly suppress the $i$ and $j$ indices, we recall $\vec s=\{s_{12},s_{23},s_{34},s_{45},s_{15}\}$, and
$r^\pm_k(\vb*{h}, \vec s)$ are rational functions of the Mandelstam variables. 
The analytic continuation of the $r_k$ from Euclidean to physical
region is therefore trivial.
For a given $i$ and $j$, the $r_k(\vb*{h},\vb*{p})$ are not
all linearly independent of each other (over $\mathbb{Q}$)  for different values of $k$, and
we can define a basis of rational functions 
$\vec v(\vb*{h},\vb*{p})=\{v_1(\vb*{h},\vb*{p}),\ldots\}$ such
that
all $r_k(\vb*{h},\vb*{p})$ live in the 
$\mathbb{Q}$-linear span of $\vec v(\vb*{h},\vb*{p})$. We can
then write
\begin{equation}\label{eq:remainderDot}
  \mathcal{R}^{(i)[j]}(\vb*{h},\vb*{p}_E)=
  \vec v^{(i)[j]}(\vb*{h},\vb*{p}_E)\cdot M^{(i)[j]}_E(\vb*{h})\cdot\vec f(\vb*{p}_E)\,,
\end{equation}
where $M_E^{(i)[j]}$ is a matrix of rational numbers.

To understand our approach to the analytic continuation of
eq.~\eqref{eq:remainderDot}, let us begin by 
considering what would happen if we had chosen a different basis of pentagon functions
$\vec f'(\vb*{p}_E)$, such that $\vec f'(\vb*{p}_E)=N\cdot \vec f(\vb*{p}_E)$ where
$N$ is a matrix of rational numbers. It is clear that to write \cref{eq:remainderDot}
in this new basis of functions we do not need to determine the
basis 
$\vec v(\vb*{h},\vb*{p}_E)$ of rational functions. The remainders
would have
exactly the same form, but with the matrix $M_E$ replaced by 
$M'^{(i)[j]}_E=M^{(i)[j]}_E\cdot N^{-1}$, which is just a different matrix of rational numbers.

Let us now discuss what happens with the pentagon functions under analytic continuation,
which we will do through an example.
Consider the simplest non-trivial pentagon function, $\log(-s_{12})$.
The choice to write it in this form follows from the fact that in the Euclidean region  $s_{12}<0$ and
the master integrals are real valued.
We note that, in this region, we could also have written $\log(\abs{s_{12}})$.
Let us now continue this function to the physical region in  \cref{eq:physregion}, where $s_{12}>0$. 
We find
\begin{equation}
  \log(-s_{12})=\log(\abs{s_{12}})-\ii\pi\,
\end{equation}
where the sign is determined by the usual Feynman $\ii\varepsilon$-prescription of 
the propagators. We see that the analytic continuation into the physical region introduces
a new pentagon ``function'', the transcendental constant $\ii\pi$. Crucially, however, the coefficient
of this function is generated by the analytic continuation of $\log(-s_{12})$. This implies
that the coefficient of $\ii\pi$ 
is trivially related to the coefficient of $\log(-s_{12})$
when analytically continuing \cref{eq:mastersToPent} or \cref{eq:remAsPent}, or more precisely that it is
in the $\mathbb{Q}$-linear span of 
$\vec v(\vb*{h},\vb*{p}_\varphi)$ (we recall that $\vb*{p}_\varphi$ is used to denote
a point in the physical region).
We assume that this
is a general feature of the analytic continuation of pentagon
functions: while the dimension of the vector space in \cref{eq:remAsPent}
might increase, the coefficients of the new pentagon functions
are not independent
from the coefficients of the pentagon functions in the Euclidean region.\footnote{
  In general, the two will be related by the (integer) winding numbers that naturally appear
  when analytically continuing multi-valued functions.
}
Even though we do not prove this statement, as we shall see below we can explicitly check that
it holds for the remainders in \cref{eq:remainderDot}.

To analytically continue the remainders computed in ref.~\cite{Abreu:2019odu}, 
we use the set of pentagon functions defined in ref.~\cite{Chicherin:2020oor}, and we denote the vector space of their monomials 
as~$\vec g(\vb*{p}_\varphi)$.
These are valid precisely in the kinematic region of \cref{eq:physregion}, and  are related to the set 
$\vec f(\vb*{p}_E)$ used in ref.~\cite{Abreu:2019odu} by a 
change of basis and by analytic continuation.
Given the points raised above
about these two operations, this means that remainders in the physical 
region can be written in terms of the same bases of rational functions that were found for
the Euclidean region. Hence we can write the partial remainder in the
physical region as
\begin{equation}\label{eq:remPhys}
  \mathcal{R}^{(i)[j]}_\varphi(\vb*{h}, \vb*{p}_\varphi; \mu)=
  \vec v^{(i)[j]}(\vb*{h}, \vb*{p}_\varphi)\cdot M^{(i)[j]}_\varphi(\vb*{h}) \cdot\vec g(\vb*{p}_\varphi/\mu)\,,
\end{equation}
where for completeness we made the dependence on $\mu$ explicit.
We stress that this representation is valid only when $p_1$ and $p_2$ are in the initial state
and $p_3$, $p_4$ and $p_5$ are in the final state, because that is the region of validity of the pentagon 
functions~$\vec g(\vb*{p}_\varphi)$. 
We discuss the generalization to
other configurations in \cref{sec:partialsAssemble}.
The $M^{(i)[j]}_{\varphi}(\vb*{h})$ can be determined from a small number of
numerical evaluations of the amplitudes. To do this, we take equation
\cref{eq:remPhys} as an ansatz for the form of the remainder when expressed in
terms of the physical region pentagon functions $\vec{g}(\vb*{p}_\varphi)$, with
the entries of the $M^{(i)[j]}_{\varphi}(\vb*{h})$ as unknowns.
We then constrain the unknowns by performing the analogous procedure of
eqs.~\eqref{eq:mastersToPent} and \eqref{eq:remAsPent}
at a sufficient number of phase-space points.\footnote{This
number is dictated by
the number of independent functions in $\vec v^{(i)[j]}(\vb*{h},\vb*{p})$.
The most complicated case, the five-gluon remainder
$\mathcal{R}^{(2)[0]}_\varphi(\{-1,+1,-1,+1,+1\},\vb*{p})$, requires 57 evaluations. 
We note that a single numerical evaluation gives more than one constraint.}
Linear algebra then allows us to extract the values of
$M^{(i)[j]}_\varphi(\vb*{h})$.

In practice, we use the implementation of two-loop numerical unitarity in
\Caravel{}~\cite{Abreu:2020xvt} to generate all the required numerical data. For
efficiency reasons, the procedure is performed over finite fields and we
construct the $\mathbb{Q}$-valued $M^{(i)[j]}_\varphi(\vb*{h})$ via the
Chinese remainder theorem and standard rational reconstruction techniques.
We remark that we used at most three finite fields of cardinality $\mathcal{O}(2^{32})$.
Following these steps, we obtain the analytic continuation of the
partial amplitudes computed in ref.~\cite{Abreu:2019odu} to the physical region
defined in \cref{eq:physregion}, corresponding to the channels
in \cref{eq:gProc,eq:qProc,eq:qQProc,eq:qQProcP,eq:4qProc}.
Crucially, this linear algebra exercise would not allow one to compute the
matrices $M^{(i)[j]}_\varphi(\vb*{h})$ in the ansatz \eqref{eq:remPhys} if the assumption on the analytic
continuation of the pentagon functions did not hold.

\subsection{From Partial Amplitudes to Amplitudes}\label{sec:partialsAssemble}

For physical applications, it is not sufficient to know a single partial
amplitude for each channel. Indeed, we must be able to 
evaluate all the terms in the sum in \cref{eq:colourDec} for each
helicity configuration. 
We start by summarizing this information for each
of the channels listed in
\cref{eq:gProc,eq:qProc,eq:qQProc,eq:qQProcP,eq:4qProc}. We remind the reader of
the comment on momenta conventions under \cref{eq:qQProcP}.

\paragraph{Five gluon:} There is a single channel, see \cref{eq:gProc}.
The set $\Sigma=S_5/ Z_5$ contains 24 elements, and the color factor
for $\sigma_0=\{1,2,3,4,5\}$ is
\begin{equation}
  \mathcal{C}(\sigma_0) \coloneqq \tr{T^{a_1}T^{a_2}T^{a_3}T^{a_4}T^{a_5}}\,.
\end{equation}
For each of the 24 partial amplitudes, there are 32 different
helicity configurations. Accounting for relations under charge conjugation
and parity, there are 192 partial helicity amplitudes to compute.
We can further reduce this set by considering the $S_2\times S_3$
permutations of initial state and final state labels which, crucially,
still correspond to the physical region defined at the start of \cref{sec:physReg}.
This leaves us with a generating set of 32 amplitudes.

\paragraph{Single quark line:} There are three channels, see \cref{eq:qProc}.
For the channel 
$\bar{q}^{h_1}_{p_1}q^{h_2}_{p_2}\to \gluon^{h_3}_{p_3}\gluon^{h_4}_{p_4}\gluon^{h_5}_{p_5}$,
the set $\Sigma=\{1,2\} \otimes S_3(3,4,5)$ contains 6 elements, and the color factor
for $\sigma_0=\{1,2,3,4,5\}$ is
\begin{equation}
  \mathcal{C}(\sigma_0) \coloneqq \qty(T^{a_3}T^{a_4}T^{a_5})^{\bar{i}_2}_{i_1}\,.
\end{equation}
For each of the three channels there are 6 partial amplitudes,
each with 16 different helicity configurations.
Accounting for relations under charge conjugation
and parity, there are 144 partial helicity amplitudes to compute.
Considering the $S_2\times S_3$
permutation of initial state and final state labels further reduces the
cardinality of the generating set to 40.

\paragraph{Two quark lines:} There are four channels, see \cref{eq:qQProc},
but we also include the two channels in \cref{eq:qQProcP} in order to consider
two identical quark lines.
For the channel 
$\bar{q}^{h_1}_{p_1}q^{h_2}_{p_2}\to Q^{h_3}_{p_3}\bar{Q}^{h_4}_{p_4}\gluon^{h_5}_{p_5}$,
the set $\Sigma=\{\{1,2,3,4,5\},\{3,4,1,2,5\}\}$ contains 2 elements, and the color factor
for $\sigma_0=\{1,2,3,4,5\}$ is
\begin{equation}
  \mathcal{C}(\sigma_0) \coloneqq \qty(T^{a_5})^{\bar{i}_4}_{i_1}
  \delta^{\bar{i}_2}_{i_3}\,.
\end{equation}
For each of the six channels there are 2 partial amplitudes,
each with 8 different helicity configurations.
Accounting for relations under charge conjugation
and parity, there are 48 partial helicity amplitudes to compute.
Considering the $S_2\times S_3$
permutation of initial state and final state labels further reduces the
cardinality of the generating set to 24.\\ %

Having classified a generating set of partial amplitudes for each channel
which respect the region constraints of the pentagon functions, we can now apply
an analogous procedure to that of \cref{sec:physReg} to each of these partial
amplitudes.
Let us consider the remainder of a partial amplitude where the helicities and
momenta are related by a permutation $\sigma$ to those in \cref{eq:remPhys}.
Clearly, the master integrals in the decomposition \eqref{eq:ampAsMaster} of the 
permuted amplitude 
are related to those of the original amplitude by the permutation $\sigma$.
We remind the reader that the pentagon functions of
ref.~\cite{Chicherin:2020oor} form a basis of the space of transcendental
functions with $p_1$ and $p_2$ initial state and $p_3$, $p_4$ and $p_5$ final state.
Hence we can express any permutation of the master integrals with this
configuration of the momenta as
\begin{equation}\label{eq:permutedMastersToPhysicalPent}
  m^{(i)}_k(\sigma(\vb*{p});\eps)=\sum_{k=0}^4\sum_{l\in B_\varphi}\epsilon^k d^{(i)}_{k,l}(\sigma) g_l(\vb*{p}) 
  + \mathcal{O}(\epsilon^5)\,,
\end{equation}
where $B_\varphi$ is the set of labels distinguishing the elements of $\vec g(\vb*{p})$.
It therefore follows that all permutations of the partial remainders 
contributing to the remainder in \cref{eq:colourDecRem}
can be expressed in the same basis of pentagon functions. That is, a
partial finite remainder evaluated with a permutation of the input helicities
$\sigma(\vb*{h})$ and on a permutation of the input point
$\sigma(\vb*{p})$ can be expressed in the form
\begin{equation}\label{eq:remPhysPermuted}
  \mathcal{R}^{(i)[j]}_\varphi(\sigma(\vb*{h}), \sigma(\vb*{p}); \mu)=
  \vec v^{(i)[j]}(\sigma(\vb*{h}), \sigma(\vb*{p}))\cdot M^{(i)[j]}_\varphi(\sigma(\vb*{h}), \sigma) \cdot\vec g(\vb*{p} / \mu)\,,
\end{equation}
where, as in \cref{eq:remPhys}, we made the dependence on $\mu$ explicit.
We recall that we can easily construct the bases $\vec v^{(i)[j]}(\sigma(\vb*{p}),\sigma(\vb*{h}))$,
as the action of $\sigma$ on rational functions is trivial to evaluate.
The matrices $M^{(i)[j]}_\varphi(\sigma(\vb*{h}), \sigma)$ can then be determined by the same ansatz procedure described
for the analytic continuation of the results of ref.~\cite{Abreu:2019odu}.
Altogether, this allows us to obtain all the expressions required to evaluate the squared
finite remainders defined in \cref{sec:sqR} for the channels in
\cref{eq:gProc,eq:qProc,eq:qQProc,eq:qQProcP,eq:4qProc}.


\section{Results and Validation}\label{sec:results}

In this section we describe our main results, which are of two types. 
The first is a collection of ancillary files 
containing analytic expressions for a generating set of (partial) remainders for each
of the channels in \cref{eq:gProc,eq:qProc,eq:qQProc,eq:qQProcP,eq:4qProc}. The second is
a \texttt{C++} library that allows for an efficient numerical evaluation of the 
finite remainders defined in \cref{eq:remPhys} for all required
permutations, as well as
the squared finite remainders defined in \cref{eq:h1,eq:h2}. 
Finally, we describe the checks we have performed on our results.

\subsection{Analytic Results}\label{sec:anaResult}

In the ancillary files we give all necessary ingredients
to assemble analytic expressions for a generating
set of two-loop partial remainders
$\mathcal{R}^{(2)[j]}_\varphi(\sigma(\vb*{h}), \sigma(\vb*{p}))$, 
for each of the channels in \cref{eq:gProc,eq:qProc,eq:qQProc,eq:qQProcP,eq:4qProc}.
From this generating set, all the
permutations required in \cref{eq:colourDecRem} can be
constructed by applying (a combination of) charge conjugation,
parity transformations,%
\footnote{ 
  Parity transformation of the analytic expressions are performed by changing the signs of all parity-odd pentagon functions and $\trFive$.
}
and the $S_2 \cross S_3$ permutations of initial and final state labels.%
\footnote{
  One may wish to express the $S_2 \cross S_3$ permutations of the partial remainders from the generating set in terms of the pentagon functions evaluated on a non-permuted phase-space point.
  This can be achieved by following the procedure described in \cref{sec:partialsAssemble}.
}

The expressions we present in the ancillary
files allow to explicitly construct the generating set of
remainders in the decomposition of \cref{eq:remPhysPermuted}.
The basis of transcendental functions 
$\vec g(\vb*{p})$ is universal
across all channels and powers of $N_f$, and is 
given in the file \texttt{anc/pentagonFunctionBasis.m}, 
written in the notation of the \Mathematica{} package 
provided with ref.~\cite{Chicherin:2020oor}.
The basis of rational functions
$\vec v^{(2)[j]}(\sigma(\vb*{h}), \sigma(\vb*{p}))$ 
is channel and
power-of-$N_f$ specific but trivially dependent on $\sigma$ (more
precisely, the action of $\sigma$ on 
$\vec v^{(2)[j]}(\sigma(\vb*{h}), \sigma(\vb*{p}))$ 
simply amounts to the permutation of the labels
in the Mandelstam invariants appearing in the rational
functions). For each channel and power of $N_f$, we 
simply reuse the bases in the ancillary files of
ref.~\cite{Abreu:2019odu} (we note again the comment below \cref{eq:qQProcP}
regarding the conventions to denote each channel). 
For completeness, we include them
in the file 
\texttt{anc/rationalBases.m} in the format
\begin{center}
  \texttt{\{channel, helicity, Nf\} -> \{list of rational functions in ($\vec{s}$, $\trFive$)\}}. 
\end{center}
We note that one can alternatively use the form of the same bases of rational
functions determined in ref.~\cite{DeLaurentis:2020qle}, which are more
compactly expressed in the spinor-helicity
formalism.
Finally, the matrices $M^{(2)[j]}_\varphi(\sigma(\vb*{h}), \sigma)$
for each member of the generating sets of remainders
are given explicitly
in \texttt{anc/twoLoopProjectionMatrices.m}.
Each is expressed in the form
\begin{center}
  \texttt{\{channel, helicity, NfPower, permutation\} -> matrix}. 
\end{center}
where \texttt{matrix} is in \Mathematica{}'s 
\texttt{SparseArray} format.

As described in \cref{sec:method}, these matrices 
constitute the main analytic result of this paper. Indeed,
they were the only missing information required to
obtain analytic expressions for the five-parton amplitudes 
in the physical region
corresponding to three-jet production at hadron colliders
given in \cref{eq:physregion}.

\subsection{Numerical Evaluation}\label{sec:numResult}

Having phenomenological applications in mind, we implemented a \texttt{C++} library
that allows for the efficient evaluation of finite remainders and squared finite remainders.
In this section we comment on this numerical implementation.

Let us start by describing the quantities which can be evaluated with our \texttt{C++} library~\cite{FivePointAmplitudes}.
The code performs the numerical evaluation of one- and two-loop (partial) remainders
for each channel, each helicity configuration and each power of $N_f$. That is, we evaluate each of the
$\mathcal{R}^{(i)[j]}\left(\sigma(\vb*{h}),\sigma(\vb*{p})\right)$ for $i=1,2$ and present each power
of $(N_f/N_c)^j$ separately. We stress that we have implemented all permutations, and not only the generating set
for which we provide analytic expressions. For physical applications, one might be interested
in the squared finite remainders defined in \cref{eq:h1,eq:h2}. Our code also outputs the numerical
values for these quantities, once again presenting each power of $N_f/N_c$ separately, that is the 
$H^{(i)[j]}$ in \cref{eq:h1NfDec,eq:h2NfDec}.

We employ \texttt{PentagonFunctions++} \cite{Chicherin:2020oor} for the numerical evaluation of the pentagon functions.
For the efficient evaluation of the rational functions, we generate optimized code with \texttt{FORM}~\cite{Ruijl:2017dtg,Kuipers:2013pba}.
In any of the helicity remainders $\mathcal{R}^{(2)[j]}\left(\sigma(\vb*{h}),\sigma(\vb*{p})\right)$ that we compute numerically,
the dominant contribution to the evaluation time is that of the pentagon functions.
When computing $H^{(1)}$ and $H^{(2)}$, we have thus organized our numerical code such that the pentagon functions are
evaluated only once per phase-space point,
independently of the number of permutations contributing to the sums in \cref{eq:h1,eq:h2}.
We note nevertheless that, compared to e.g.~the case of triphoton production \cite{Abreu:2020cwb}, 
the evaluation of $H^{(2)}$ for five-gluon channels receives contributions from a large number of partial amplitudes, and therefore
the time spent in the evaluation of the other components in \cref{eq:remPhysPermuted} becomes noticeable.
We find that the typical double-precision evaluation time for any $H^{(2)}$ is $\mathcal{O}(1\text{s})$ on a single CPU, 
while the evaluation time for $H^{(1)}$ is negligible at $\mathcal{O}(10\text{ms})$.

To be suitable for phenomenological applications the numerical evaluations should not only be fast but also stable across phase space.
To guarantee the latter, we develop a precision-rescue strategy that detects and corrects unstable numerical evaluations.
The main source of numerical instabilities is the presence of large cancellations between different contributions in \cref{eq:remPhysPermuted}, when
a phase-space point gets close to the zero sets of the rational functions' denominators.
As we have already argued in refs.~\cite{Abreu:2018zmy,Abreu:2019odu,Abreu:2020cwb},
these are determined by a set of 25 quantities $\{W_i(\vec s)\}$, with $i=1,\ldots,25$, which are 
a subset of the so-called {alphabet} associated with the five-point
two-loop master integrals (see e.g.~ref.~\cite{Chicherin:2017dob} for the definition of these quantities).
This subset is rational in the Mandelstam variables $\vec s$ defined in \cref{eq:mandDef} and all 
elements have the same dimensions.
We find that potentially unstable points can be efficiently identified by computing a quantity that characterizes the spread of scales for a given phase-space point $\vec s$,\footnote{%
  We attempted to further reduce the subset of letters to only include those that correspond to spurious singularities, as these had been identified
  in ref.~\cite{Abreu:2020cwb} as being sufficient to characterize the points that were less stable. We find that for
  the five-parton channels considered in this paper this subset covers most of the unstable points, but we must also include non-spurious singularities to detect all unstable points.
}
\begin{equation}\label{eq:letters-check}
  \kappa(\vec s) \coloneqq \frac{\min_i\{\abs{W_i(\vec s)}\}}{\max_i\{\abs{W_i(\vec s)}\}}\,.
\end{equation}
We introduce a threshold $\kappa_\text{thr}$,
and if $\kappa(\vec s)>\kappa_\text{thr}$ the point is considered stable.
Otherwise, the point is potentially unstable and 
we perform a second evaluation on an infinitesimally perturbed point $\vec s_{\delta}$ defined as
\begin{equation}
  \vec s_{\delta} - \vec s \simeq {\varepsilon_\text{double}} \, \vec s\,,
\end{equation}
where $\varepsilon_\text{double} \simeq 10^{-16}$ is the machine epsilon of double-precision floating point numbers.\footnote{%
In practice, each of the components $(\vec s_{\delta})_i$ is taken to be the next (or previous) representable floating-point number after $(\vec s)_i$.
}
We estimate the accuracy of potentially unstable points as 
\begin{equation}\label{eq:accuracy-estimate}
  \Delta(\vec s) \coloneqq  \abs{ 1 - \frac{H(\vec{s}_\delta)}{H(\vec{s})}}\,.
\end{equation}
We introduce another threshold value $\Delta_\text{thr}$, and if $\Delta(\vec s) < \Delta_\text{thr}$ the point is considered
stable. If $\Delta(\vec s) > \Delta_\text{thr}$, the point $\vec s$ is considered unstable and evaluated in quadruple precision.
The performance of this rescue strategy and the values we choose for $\kappa_\text{thr}$ and 
$\Delta_\text{thr}$ will be discussed below.

\begin{figure}[t]
  \centering
  \begin{subfigure}[t]{0.68\linewidth}
    \includegraphics[width=1\linewidth]{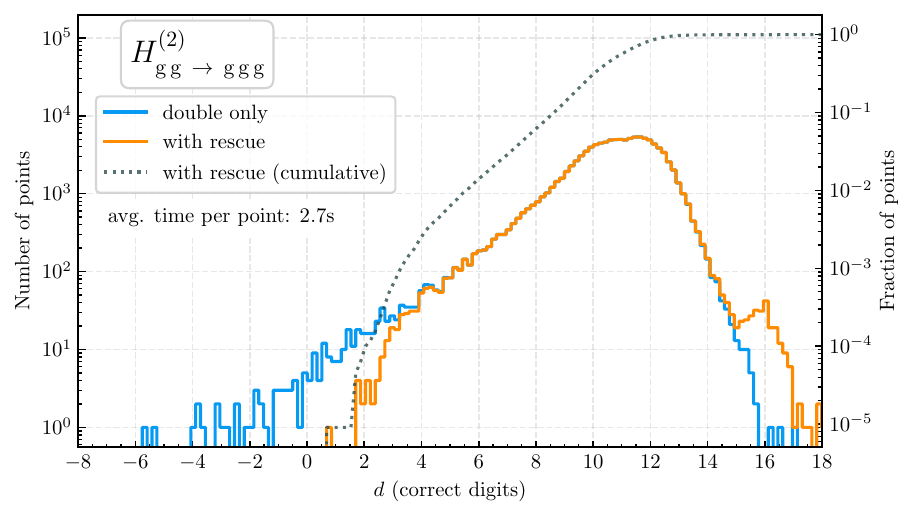}
  \end{subfigure}
  \begin{subfigure}[t]{0.68\linewidth}
    \includegraphics[width=1\linewidth]{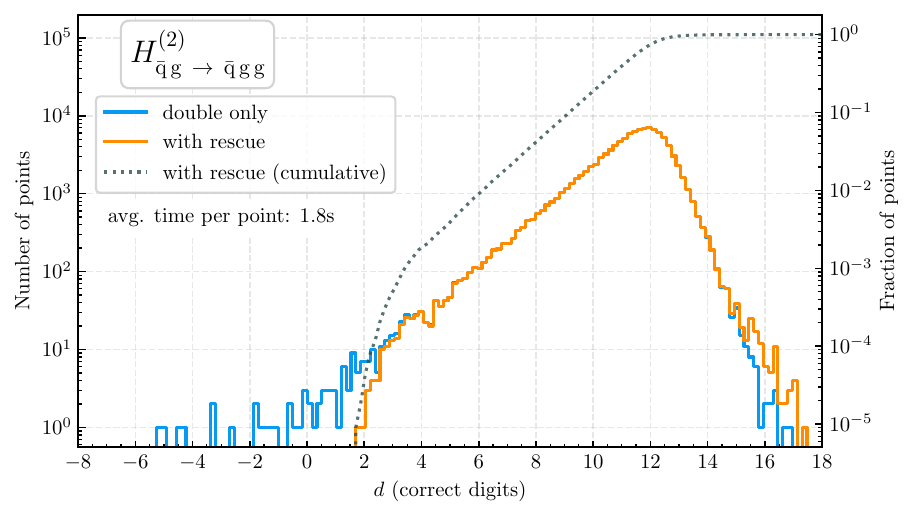}
  \end{subfigure}
  \begin{subfigure}[t]{0.68\linewidth}
    \includegraphics[width=1\linewidth]{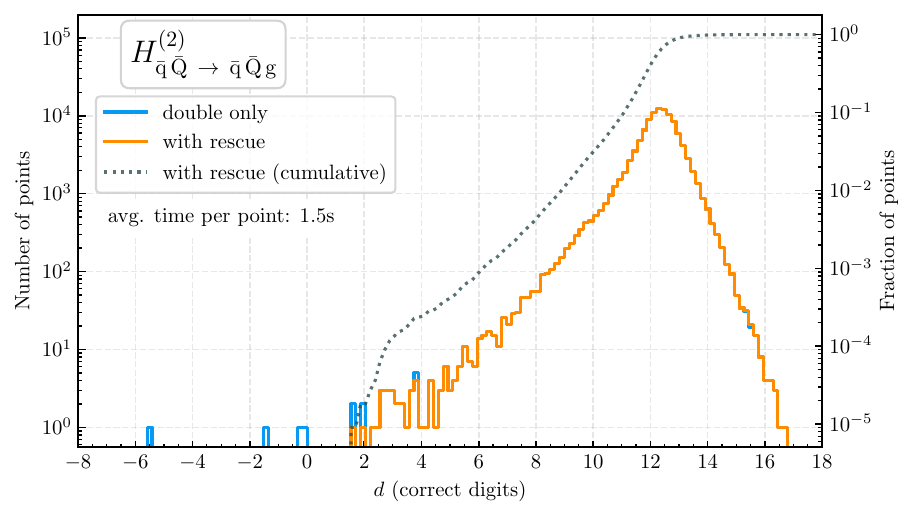}
  \end{subfigure}
  \caption{Logarithmic distribution of correct digits (see \cref{eq:correct-digits}) on samples of 110k points for each of the channels.
    The blue line corresponds to evaluation in double precision. The orange line represents evaluation with the rescue system enabled.
    The gray dashed line is the cumulative distribution of the latter. The rescue system's thresholds are the same as in \cref{tab:rescue-data}.
    Average core time for evaluation on a single phase-space point with rescue
    system enabled as measured when evaluating all sample points in parallel using all 32 threads of \texttt{Intel(R) Xeon(R) Silver 4216 CPU @ 2.10GHz}.
  }
  \label{fig:numerical_stability}
\end{figure}

To demonstrate the performance of our numerical \texttt{C++} implementation, we employ \textsc{Sherpa} 2.2 \cite{Bothmann:2019yzt} to 
sample phase-space points from a Monte-Carlo integration grid optimized on the Born matrix elements for three-jet production.
We sample a sufficient number of points to draw 110k events for each of the channels $\gluon\gluon\to \gluon\gluon\gluon$, $\bar{q}\gluon\to \bar{q}\gluon\gluon$ and $\bar{q}q\to Q\bar{Q}\gluon$,
which are representative of all channels in \cref{eq:gProc,eq:qProc,eq:qQProc}.
We use the same phase-space definition that was used in the 7 TeV \textsf{ATLAS} analysis \cite{Aad:2014rma}, which we summarize here for completeness.
We require three hard jets defined with the anti-$k_t$ algorithm,
with the radius parameter of $R=0.4$, as implemented in \textsc{FastJet} \cite{Cacciari:2011ma}.
We require the jets to be within the rapidity range $\abs{y} < 3$ and that they all have transverse momenta larger than $50$~GeV. 
In addition, the leading and subleading jets are required  to have transverse momenta larger than $150$~GeV and  $100$~GeV respectively.
We then evaluate $H^{(2)}$ at the dynamical scale
\begin{equation}
  \mu  = \frac{1}{2} \sum_{j=1}^{3} p_{Tj}\,,
\end{equation}
and sum the contributions of different powers of $\NF/\NC$, setting $\NC=3$ and $\NF=5$.
The accuracy of double-precision evaluations is determined by comparing them to an evaluation in quadruple precision as
\begin{equation} \label{eq:correct-digits}
  d \coloneqq -\log_{10}\abs{ 1 - \frac{H^{(2)}_\text{double}}{H^{(2)}_\text{quad} }}\,,
\end{equation}
where we keep the dependence on the phase-space point implicit.
If the rescue system is turned on, then the accuracy of the rescued points that were evaluated in quadruple
precision is estimated with a second quadruple-precision evaluation as in \cref{eq:accuracy-estimate}.
In \cref{fig:numerical_stability} we present the distribution of correct digits for the three representative channels,
obtained over the 110k phase-space points. We plot the curve with the rescue system turned off (blue lines), which shows
that a rescue system is indeed required for a few points. In orange we plot the curve with the rescue system turned on,
with $\kappa_\text{thr} = 10^{-4}$ and $\Delta_\text{thr} = 10^{-2}$,
and we see that all unstable points have been correctly rescued and the
numerical stability is adequate. 
The gray dashed line is the cumulative distribution with the rescue system enabled.
On the plots, we also overlay the average core time for a single phase-space point evaluation with the rescue system turned on.

To assess the impact of the rescue system on the timings of the numerical evaluation, we collected in \cref{tab:rescue-data}
some characterizing data for the choice of thresholds $\kappa_\text{thr} = 10^{-4}$ and $\Delta_\text{thr} = 10^{-2}$.
The test in \cref{eq:letters-check} catches too many points, but crucially catches all unstable points as can be seen
from \cref{fig:numerical_stability}.
The second test, in \cref{eq:accuracy-estimate}, is much more precise but also more expensive in terms of evaluation
time. Overall, we see that our rescue system catches all necessary points with only a mild slowdown factor.
The values $\kappa_\text{thr} = 10^{-4}$ and $\Delta_\text{thr} = 10^{-2}$ can be modified by the user according to their needs.
It would be interesting to investigate if using the different form of the bases of rational functions derived in 
ref.~\cite{DeLaurentis:2020qle} reduces the fraction of unstable points, but we find that our implementation is already
satisfactorily fast and stable for phenomenological applications.

\begin{table}[ht]
  \renewcommand{\arraystretch}{1.4}
  \newcolumntype{T}{>{\centering\arraybackslash}m{13ex}}
  \centering
  \begin{tabular}{cTTTTT}
    \toprule
    Channels & \% recomputed in double & \% recomputed in quad & slowdown factor \\
    \midrule
    $H^{(2)}_{\mathrm{\gluon}\,\mathrm{\gluon} ~\to~ \mathrm{\gluon}\,\mathrm{\gluon}\,\mathrm{\gluon}}$ &  2.52 & 0.27  &  1.12 \\
    $H^{(2)}_{\bar{\mathrm{q}}\,\mathrm{\gluon} ~\to~ \bar{\mathrm{q}}\,\mathrm{\gluon}\,\mathrm{\gluon}}$ &  2.57 & 0.08  &  1.06 \\
    $H^{(2)}_{\bar{\mathrm{q}}\,\bar{\mathrm{Q}} ~\to~ \bar{\mathrm{q}}\,\bar{\mathrm{Q}}\,\mathrm{\gluon}}$ &  2.52 & 0.01  &  1.03 \\
    \bottomrule
  \end{tabular}
  \caption{
    Characteristic performance of the rescue system on the samples of 110k points with $\kappa_\text{thr} = 10^{-4}$ and $\Delta_\text{thr} = 10^{-2}$.
    The first column shows the fraction of points with $\kappa(\vec s)<\kappa_\text{thr}$, which are recomputed in double precision 
    at $\vec s_\delta$.
    The second column shows the fraction of points with $\Delta(\vec s) > \Delta_\text{thr}$, which are recomputed in 
    quadruple precision at $\vec s$.
    The average slowdown due to the usage of the rescue system w.r.t. only double-precision evaluations is shown in the last column.
  }
  \label{tab:rescue-data}
\end{table}

In order to facilitate the comparison with our results, we present in \cref{sec:target}
the numerical values of squared finite remainders at each power of $\NF$,
see \cref{eq:h1NfDec,eq:h2NfDec}, at a randomly chosen phase-space point.

\subsection{Validation}

The main analytic results of our paper are the matrices 
described in \cref{sec:anaResult}. These are obtained by 
building on the analytic results of ref.~\cite{Abreu:2019odu},
which have undergone very stringent checks. Compared to
ref.~\cite{Abreu:2019odu}, our current results are written in terms of 
a different basis of transcendental functions \cite{Chicherin:2020oor}
which have also undergone several direct checks (e.g., comparison
with the functions of ref.~\cite{Gehrmann:2018yef}) 
and indirect checks \cite{Kallweit:2020gcp,Abreu:2020cwb,
Chawdhry:2020for,Agarwal:2021grm}. Furthermore, we have verified
that we reproduce the results independently obtained in 
ref.~\cite{Badger:2018gip}
for the numerical evaluation of five-gluon amplitudes in the
physical region defined in \cref{eq:physregion}.
Finally, we compared our results for the function $H^{(2)[0]}_{\gluon\gluon\to \gluon\gluon\gluon}$ in the purely gluonic channel against an independent computation \cite{Badger:correspondence} and found agreement.

We have also performed several consistency checks on our numerical results.
Since we obtain all the helicity partial remainders required for the functions $H^{(1)}$ and $H^{(2)}$
from sums over the permutations of the generating sets of remainders introduced in \cref{sec:partialsAssemble},
it is important to check that these operations
are correctly implemented in the \texttt{C++} library we have developed.
We have made three types of consistency checks to validate this procedure.

\paragraph{Symmetries.} We have checked that the remainders and squared remainders satisfy all expected charge, parity conjugation and permutation symmetries.

\paragraph{Validation of $H^{(1)}$.} We have compared the one-loop squared remainders $H^{(1)}$ of all the channels in \cref{eq:gProc,eq:qProc,eq:qQProc,eq:qQProcP,eq:4qProc}
against the evaluations by the \texttt{BlackHat} library \cite{Berger:2008sj} and found full agreement.%
\footnote{
  After adjusting the definitions of the leading-color squared matrix elements in \texttt{BlackHat} such that all subleading-color contributions are discarded.
}
The color-dressed remainders $\mathcal{R}^{(1)}_\mathcal{M}$ in \cref{eq:colourDecRem} are constructed from generating sets of partial
remainders, as described in \cref{sec:partialsAssemble}, and then are directly fed into the definition of $H^{(1)}$, see \cref{eq:h1}.
The procedure for assembling the squared remainders is independent of the number of loops, hence
this check validates the corresponding assembly of the contributions for the calculation of $H^{(2)}$.

\paragraph{Scale dependence of remainders.} 
Given that we evaluate the finite remainders directly, the correct pole structure of the underlying scattering amplitudes
is satisfied by construction. We note nevertheless that
finite remainders retain a memory of the poles of the 
amplitudes and of their explicit (scheme-dependent) definition.
This information is encoded in the scale dependence of the finite remainders. 
Quite generally, we find that
\begin{align}\begin{split}\label{eq:remScaleDep}
  \frac{\partial\,\mathcal{R}^{(1)}(\mu)}{\partial \log \mu}~=~&
  \frac{2 \beta_0\lambda}{N_c}\mathcal{R}^{(0)}\,,\\
  \frac{\partial\,\mathcal{R}^{(2)}(\mu)}{\partial \log \mu}~=~& 2 \left(\frac{\beta _1 \lambda }{N_c^2}+{\bf H}_{[n]}\right)\mathcal{R}^{(0)}
  +\frac{2\beta_0\left(2+\lambda\right)}{N_c}\mathcal{R}^{(1)}(\mu)\,,
\end{split}\end{align}
where we denote by $\lambda$ the power of $g_s^0$ in the tree-level amplitude (for five-parton amplitudes $\lambda=3$, see \cref{eq:pertExp}) and
the $\beta_i$ are the coefficients of the QCD $\beta$ function explicitly given in \cref{eq:betai}. ${\bf H}_{[n]}$ 
is given explicitly in \cref{eq:hdef,eq:hexp}
and controls the collinear singularities of scattering amplitudes. We have
verified that the scale dependence of the remainders obtained from our numerical implementation is described by \cref{eq:remScaleDep}. Furthermore, we have
independently checked that, starting from this
scale dependence, we reproduce the scale dependence of the functions $H^{(1)}$ and $H^{(2)}$, which again validates the correct assembly of these
functions.

\FloatBarrier


\section{Conclusion}\label{sec:conclusion}

In this paper we present the NNLO double-virtual contributions to three-jet production at hadron colliders
at leading color. While all required amplitudes were already available in the literature in analytic form,
they were only available in the non-physical Euclidean region, and as such could not be directly
used for phenomenological applications. Aside from analytically continuing those expressions
to the physical region, we also provide a  \texttt{C++} library that allows to efficiently
evaluate these contributions and obtain stable results across phase space.

The analytic continuation of the results of ref.~\cite{Abreu:2019odu} to the physical
region corresponding to three-jet production  is performed
by rewriting them in terms of the pentagon functions defined in ref.~\cite{Chicherin:2020oor}.
By noticing that this rewriting amounts to computing a matrix of rational numbers,
we obtained the physical-region expressions from a few numerical evaluations of the amplitudes 
(57 in the most complicated case). These evaluations were performed with the 
implementation of the two-loop numerical unitarity approach in \Caravel{} \cite{Abreu:2020xvt}.
We provide analytic expressions for a generating set of partial remainders. 
For all channels in \cref{eq:gProc,eq:qProc,eq:qQProc,eq:qQProcP,eq:4qProc},
these allow one to evaluate the color-dressed remainders defined in \cref{eq:colourDecRem} as well as
the squared remainders $H^{(1)}$ and $H^{(2)}$ defined in \cref{eq:h1NfDec,eq:h2NfDec}
in the physical region.

Our driving motivation was to provide results that are ready to be used for phenomenological
applications. To this end, we developed a \texttt{C++} library which
efficiently evaluates individual partial remainders as well as the one- and
two-loop squared remainders. The numerical evaluations can be performed in double or quadruple
precision. We find that double precision is sufficient for the vast majority of phase-space points, 
and have implemented a rescue system which detects unstable points
and reevaluates them in quadruple precision. With the rescue system enabled, we find
very good overall numerical stability and our evaluation time for the two-loop squared remainders is
$\mathcal{O}(1\text{s})$. This shows that our results and the numerical library
we provide are ready to be used in phenomenological studies.

\section*{Acknowledgments}
We thank Simon Badger and Simone Zoia for the correspondence on an independent calculation of the finite remainders in the purely gluonic channel. 
The work of B.P~is supported by the French Agence Nationale pour la Recherche, 
under grant ANR–17–CE31–0001–01.
V.S.~is supported by the European Research Council (ERC) under the European 
Union's Horizon 2020 research and innovation programme,
\textit{Novel structures in scattering amplitudes} (grant agreement No.\ 725110).
The work of F.~F.~C.\ is supported in part by the U.S. Department of Energy under 
grant DE-SC0010102.

\appendix

\section{Infrared Subtraction}\label{sec:appIR}

The divergences of renormalized two-loop amplitudes obey a universal structure \cite{Catani:1998bh,Becher:2009cu,Gardi:2009qi},
\begin{align}\begin{split}
    \CA_R^{(1)}&={\bf I}^{(1)}_{[n]}(\epsilon)
    \CA_R^{(0)}+\mathcal{O}
    (\epsilon^0)\,,\\
    \CA_R^{(2)}&={\bf I}^{(2)}_{[n]}(\epsilon)A_R^{(0)}+{\bf I}^{(1)}_{[n]}(\epsilon)
    \CA_R^{(1)}+\mathcal{O}(\epsilon^0)\,,
\end{split}\end{align}
with the renormalized amplitudes $\CA_R^{(i)}$ related to the bare amplitudes
$\CA^{(i)}$ as in \cref{eq:twoLoopUnRenorm}.
For amplitudes in the leading-color approximation the operators
$\mathbf{I}^{(1)}_{[n]}$ and $\mathbf{I}^{(2)}_{[n]}$ are 
diagonal in color space. 
The operator $\mathbf{I}^{(1)}_{[n]}$ is given by
\begin{equation}
  {\bf I}^{(1)}_{[n]}(\epsilon)=
  -\frac{e^{\gamma_E\epsilon}}{\Gamma(1-\epsilon)}
  \sum_{i=1}^n\gamma_{a_i,a_{i+1}}
  \left( -\frac{s_{i,i+1}}{\mu^2} -\ii \varepsilon\right)^{-\epsilon}\,,
\end{equation}
where $s_{i,j}=(p_i+p_j)^2$, with the indices defined cyclically,
and we made explicit the $\ii \varepsilon$ associated with Mandelstam
variables that follows from the Feynman $\ii \varepsilon$-prescription
of the propagators.
The index $a_i$ denotes a type of particle,
i.e.~$a_i\in\{\gluon,q,\bar q, Q, \bar Q\}$. 
The symbols $\gamma_{a,b}$ are symmetric
under exchange of indices, 
$\gamma_{a,b}=\gamma_{b,a}$, and given by:
\begin{align}\begin{split}
  \gamma_{\gluon,\gluon}&=\frac{1}{\epsilon^2}+
  \frac{1}{\epsilon}
  \frac{\beta_0}{\NC}\,, \qquad
  \gamma_{q,Q}=\gamma_{q,\bar Q}=
  \gamma_{\bar q, Q}=\gamma_{\bar q, \bar Q} 
  =\frac{1}{\epsilon^2}+\frac{3}{2\epsilon}\,,\\
  \gamma_{\gluon,q}&=\gamma_{\gluon,\bar q}=
  \gamma_{\gluon,Q}=\gamma_{\gluon,\bar Q}=
  \frac{\gamma_{\gluon,\gluon}+\gamma_{q,Q}}{2}\,,\qquad
  \gamma_{q,\bar q}=\gamma_{Q,\bar Q}=0\,.
\end{split}\end{align}
The operator~${\bf I}^{(2)}_{[n]}$ is
\begin{align}\begin{split} \label{eqn:Iop}
    {\bf I}^{(2)}_{[n]}(\epsilon)=&
  -\frac{1}{2}{\bf I}^{(1)}_{[n]}(\epsilon)
  {\bf I}^{(1)}_{[n]}(\epsilon)
  -\frac{2\beta_0}{\NC\epsilon}{\bf I}^{(1)}_{[n]}(\epsilon) + 
  \frac{e^{-\gamma_E\epsilon}\Gamma(1-2\epsilon)}
  {\Gamma(1-\epsilon)}
  \left(\frac{2\beta_0}{\NC\epsilon}+K\right)
  {\bf I}^{(1)}_{[n]}(2\epsilon) \\
  &+ \frac{e^{\gamma_E\epsilon}}{\epsilon\Gamma(1-\epsilon)}
  {\bf H}_{[n]}\,,
\end{split}\end{align}
where 
\begin{equation}
K=\frac{67}{9}-\frac{\pi ^2}{3}-\frac{10}{9}\frac{\NF}{\NC}\,,
\end{equation}
and ${\bf H}_{[n]}$ is a diagonal operator at 
leading color that depends on the number of external quarks 
and gluons,
\begin{align}\begin{split}\label{eq:hdef}
  {\bf H}_{[n]}&=  
  \sum_{i=1}^n\left(
  \delta_{a_i,\gluon}H_\gluon +
  (\delta_{a_i,q}+\delta_{a_i,\bar q}
  +\delta_{a_i,Q}+\delta_{a_i,\bar Q})
  H_q
  \right)\,,
\end{split}\end{align}
with
\begin{align}\begin{split}\label{eq:hexp}
  H_\gluon &= \left(\frac{\zeta_3}{2}+\frac{5}{12}+
  \frac{11\pi^2}{144}\right)
  -\left(\frac{\pi^2}{72}+\frac{89}{108}\right)\frac{N_f}{\NC}
  +\frac{5}{27}\left(\frac{N_f}{\NC}\right)^2\,,\\
  H_q &=
  \left(\frac{7\zeta_3}{4}+\frac{409}{864}
  -\frac{11\pi^2}{96}\right)
  +\left(\frac{\pi^2}{48}-\frac{25}{216}\right)\frac{N_f}{\NC}\,.
\end{split}\end{align}


\section{Reference Evaluations of Squared Finite Remainders}\label{sec:target}

\begin{table}[t]
  \renewcommand{\arraystretch}{1.4}
  \centering
  \begin{adjustbox}{width=1\textwidth}
  \begin{tabular}{cccccc}
    \toprule
    Channel&$H^{(1)[0]}$&$H^{(1)[1]}$&$H^{(2)[0]}$&$H^{(2)[1]}$&$H^{(2)[2]}$ \\
    \midrule
	$\gluon\gluon\rightarrow \gluon\gluon\gluon$ & $16.135254222$ & $0.19163044752$ & $464.47846208$ & $-58.116292408$ & $0.60077232705$ \\ 
	\midrule
	${\bar q}q\rightarrow \gluon\gluon\gluon$ & $9.5879406141$ & $-3.0604943308$ & $184.44415807$ & $-61.765802987$ & $6.3615768297$ \\ 
	${\bar q}\gluon\rightarrow {\bar q}\gluon\gluon$ & $26.908169290$ & $-3.6373308269$ & $867.25232363$ & $-230.76277359$ & $12.598811302$ \\ 
	$\gluon\gluon\rightarrow q{\bar q}\gluon$ & $24.495592766$ & $-2.5939909248$ & $745.87682394$ & $-166.84486839$ & $6.1899943330$ \\ 
	\midrule
	${\bar q}q\rightarrow Q{\bar Q}\gluon$ & $10.460907919$ & $-4.2060557725$ & $212.42454564$ & $-80.136400792$ & $8.2094005806$ \\ 
	${\bar q}Q\rightarrow Q{\bar q}\gluon$ & $27.104747640$ & $-4.0829938180$ & $705.58902507$ & $-209.42216177$ & $12.483148067$ \\ 
	${\bar q}{\bar Q}\rightarrow {\bar q}{\bar Q}\gluon$ & $42.313652168$ & $-8.0064067852$ & $1628.2933493$ & $-562.78735847$ & $44.198947852$ \\ 
	${\bar q}\gluon\rightarrow {\bar q}Q{\bar Q}$ & $28.068256507$ & $-6.3593609865$ & $935.81439233$ & $-324.32790785$ & $29.070926975$ \\ 
	\midrule
	${\bar q}q\rightarrow q{\bar q}\gluon$ & $20.846053179$ & $-4.1292696285$ & $520.14108472$ & $-160.80597165$ & $10.876062192$ \\ 
	${\bar q}{\bar q}\rightarrow {\bar q}{\bar q}\gluon$ & $42.259655399$ & $-7.9918854619$ & $1624.7163564$ & $-561.33769564$ & $44.056509019$ \\ 
	${\bar q}\gluon\rightarrow {\bar q}q{\bar q}$ & $28.497167934$ & $-6.2611415380$ & $947.84964732$ & $-322.54996102$ & $28.093290494$ \\ 
    \bottomrule
  \end{tabular}
  \end{adjustbox}
  \caption{
   Reference values for the evaluation of squared finite remainders at each power of $\NF$, as defined in \cref{eq:h1NfDec,eq:h2NfDec}
   on the phase-space point given in \cref{eq:referencePointBis}.
  }
  \label{tab:target}
\end{table}
In order to facilitate the comparison with our results,
we present in \cref{tab:target} reference values for the evaluation of
squared finite remainders at each power of $\NF$, as defined in \cref{eq:h1NfDec,eq:h2NfDec},
at the randomly chosen phase-space point
\begin{align}
  \begin{split}
    s_{12} &=  1.322500000, \quad \,\,
    s_{23} =  -0.994109498, \quad \,\,
    s_{34} =   0.264471591, \\
    s_{45} &=  0.267126049, \quad \,\,
    s_{15} =  -0.883795230, \quad \,\,
    \mu  =  1.
  \end{split}
  \label{eq:referencePointBis}
\end{align}
The results are obtained with our \texttt{C++} library.

\bibliography{main.bib}

\providecommand{\href}[2]{#2}\begingroup\raggedright\begin{thebibliography}{10}

\bibitem{Abazov:2012lua}
{\scshape D0} collaboration, \emph{{Measurement of angular correlations of jets
  at $\sqrt{s}=1.96$ TeV and determination of the strong coupling at high
  momentum transfers}},
  \href{https://doi.org/10.1016/j.physletb.2012.10.003}{\emph{Phys. Lett. B}
  {\bfseries 718} (2012) 56} [\href{https://arxiv.org/abs/1207.4957}{{\ttfamily
  1207.4957}}].

\bibitem{Chatrchyan:2013txa}
{\scshape CMS} collaboration, \emph{{Measurement of the Ratio of the Inclusive
  3-Jet Cross Section to the Inclusive 2-Jet Cross Section in pp Collisions at
  $\sqrt{s}$ = 7 TeV and First Determination of the Strong Coupling Constant in
  the TeV Range}},
  \href{https://doi.org/10.1140/epjc/s10052-013-2604-6}{\emph{Eur. Phys. J. C}
  {\bfseries 73} (2013) 2604}
  [\href{https://arxiv.org/abs/1304.7498}{{\ttfamily 1304.7498}}].

\bibitem{CMS:2014mna}
{\scshape CMS} collaboration, \emph{{Measurement of the inclusive 3-jet
  production differential cross section in proton\textendash{}proton collisions
  at 7 TeV and determination of the strong coupling constant in the TeV
  range}}, \href{https://doi.org/10.1140/epjc/s10052-015-3376-y}{\emph{Eur.
  Phys. J. C} {\bfseries 75} (2015) 186}
  [\href{https://arxiv.org/abs/1412.1633}{{\ttfamily 1412.1633}}].

\bibitem{ATLAS:2015yaa}
{\scshape ATLAS} collaboration, \emph{{Measurement of transverse energy-energy
  correlations in multi-jet events in $pp$ collisions at $\sqrt{s} = 7$ TeV
  using the ATLAS detector and determination of the strong coupling constant
  $\alpha_{\mathrm{s}}(m_Z)$}},
  \href{https://doi.org/10.1016/j.physletb.2015.09.050}{\emph{Phys. Lett. B}
  {\bfseries 750} (2015) 427}
  [\href{https://arxiv.org/abs/1508.01579}{{\ttfamily 1508.01579}}].

\bibitem{Ridder:2013mf}
A.~Gehrmann-De~Ridder, T.~Gehrmann, E.W.N.~Glover and J.~Pires, \emph{{Second
  order QCD corrections to jet production at hadron colliders: the all-gluon
  contribution}},
  \href{https://doi.org/10.1103/PhysRevLett.110.162003}{\emph{Phys. Rev. Lett.}
  {\bfseries 110} (2013) 162003}
  [\href{https://arxiv.org/abs/1301.7310}{{\ttfamily 1301.7310}}].

\bibitem{Currie:2016bfm}
J.~Currie, E.W.N.~Glover and J.~Pires, \emph{{Next-to-Next-to Leading Order QCD
  Predictions for Single Jet Inclusive Production at the LHC}},
  \href{https://doi.org/10.1103/PhysRevLett.118.072002}{\emph{Phys. Rev. Lett.}
  {\bfseries 118} (2017) 072002}
  [\href{https://arxiv.org/abs/1611.01460}{{\ttfamily 1611.01460}}].

\bibitem{Currie:2017eqf}
J.~Currie, A.~Gehrmann-De~Ridder, T.~Gehrmann, E.W.N.~Glover, A.~Huss and
  J.~Pires, \emph{{Precise predictions for dijet production at the LHC}},
  \href{https://doi.org/10.1103/PhysRevLett.119.152001}{\emph{Phys. Rev. Lett.}
  {\bfseries 119} (2017) 152001}
  [\href{https://arxiv.org/abs/1705.10271}{{\ttfamily 1705.10271}}].

\bibitem{Czakon:2019tmo}
M.~Czakon, A.~van Hameren, A.~Mitov and R.~Poncelet, \emph{{Single-jet
  inclusive rates with exact color at $ \mathcal{O} $ ($ {\alpha}_s^4 $)}},
  \href{https://doi.org/10.1007/JHEP10(2019)262}{\emph{JHEP} {\bfseries 10}
  (2019) 262} [\href{https://arxiv.org/abs/1907.12911}{{\ttfamily
  1907.12911}}].

\bibitem{Currie:2013dwa}
J.~Currie, A.~Gehrmann-De~Ridder, E.W.N.~Glover and J.~Pires, \emph{{NNLO QCD
  corrections to jet production at hadron colliders from gluon scattering}},
  \href{https://doi.org/10.1007/JHEP01(2014)110}{\emph{JHEP} {\bfseries 01}
  (2014) 110} [\href{https://arxiv.org/abs/1310.3993}{{\ttfamily 1310.3993}}].

\bibitem{Nagy:2001fj}
Z.~Nagy, \emph{{Three jet cross-sections in hadron hadron collisions at
  next-to-leading order}},
  \href{https://doi.org/10.1103/PhysRevLett.88.122003}{\emph{Phys. Rev. Lett.}
  {\bfseries 88} (2002) 122003}
  [\href{https://arxiv.org/abs/hep-ph/0110315}{{\ttfamily hep-ph/0110315}}].

\bibitem{Nagy:2003tz}
Z.~Nagy, \emph{{Next-to-leading order calculation of three jet observables in
  hadron hadron collision}},
  \href{https://doi.org/10.1103/PhysRevD.68.094002}{\emph{Phys. Rev. D}
  {\bfseries 68} (2003) 094002}
  [\href{https://arxiv.org/abs/hep-ph/0307268}{{\ttfamily hep-ph/0307268}}].

\bibitem{Kilgore:1996sq}
W.B.~Kilgore and W.T.~Giele, \emph{{Next-to-leading order gluonic three jet
  production at hadron colliders}},
  \href{https://doi.org/10.1103/PhysRevD.55.7183}{\emph{Phys. Rev. D}
  {\bfseries 55} (1997) 7183}
  [\href{https://arxiv.org/abs/hep-ph/9610433}{{\ttfamily hep-ph/9610433}}].

\bibitem{Kilgore:2000dr}
W.B.~Kilgore and W.T.~Giele, \emph{{A Next-to-leading order calculation of
  hadronic three jet production}},  in \emph{{30th International Conference on
  High-Energy Physics}}, pp.~502--505, 7, 2000
  [\href{https://arxiv.org/abs/hep-ph/0009193}{{\ttfamily hep-ph/0009193}}].

\bibitem{Frederix:2018nkq}
R.~Frederix, S.~Frixione, V.~Hirschi, D.~Pagani, H.S.~Shao and M.~Zaro,
  \emph{{The automation of next-to-leading order electroweak calculations}},
  \href{https://doi.org/10.1007/JHEP07(2018)185}{\emph{JHEP} {\bfseries 07}
  (2018) 185} [\href{https://arxiv.org/abs/1804.10017}{{\ttfamily
  1804.10017}}].

\bibitem{Reyer:2019obz}
M.~Reyer, M.~Sch\"onherr and S.~Schumann, \emph{{Full NLO corrections to 3-jet
  production and $\mathbf {R_{32}}$ at the LHC}},
  \href{https://doi.org/10.1140/epjc/s10052-019-6815-3}{\emph{Eur. Phys. J. C}
  {\bfseries 79} (2019) 321}
  [\href{https://arxiv.org/abs/1902.01763}{{\ttfamily 1902.01763}}].

\bibitem{Badger:2013gxa}
S.~Badger, H.~Frellesvig and Y.~Zhang, \emph{{A Two-Loop Five-Gluon Helicity
  Amplitude in QCD}},
  \href{https://doi.org/10.1007/JHEP12(2013)045}{\emph{JHEP} {\bfseries 12}
  (2013) 045} [\href{https://arxiv.org/abs/1310.1051}{{\ttfamily 1310.1051}}].

\bibitem{Badger:2017jhb}
S.~Badger, C.~Br\o{}nnum-Hansen, H.B.~Hartanto and T.~Peraro, \emph{{First look
  at two-loop five-gluon scattering in QCD}},
  \href{https://doi.org/10.1103/PhysRevLett.120.092001}{\emph{Phys. Rev. Lett.}
  {\bfseries 120} (2018) 092001}
  [\href{https://arxiv.org/abs/1712.02229}{{\ttfamily 1712.02229}}].

\bibitem{Abreu:2017hqn}
S.~Abreu, F.~Febres~Cordero, H.~Ita, B.~Page and M.~Zeng, \emph{{Planar
  Two-Loop Five-Gluon Amplitudes from Numerical Unitarity}},
  \href{https://doi.org/10.1103/PhysRevD.97.116014}{\emph{Phys. Rev. D}
  {\bfseries 97} (2018) 116014}
  [\href{https://arxiv.org/abs/1712.03946}{{\ttfamily 1712.03946}}].

\bibitem{Badger:2018gip}
S.~Badger, C.~Br\o{}nnum-Hansen, T.~Gehrmann, H.B.~Hartanto, J.~Henn,
  N.A.~Lo~Presti et~al., \emph{{Applications of integrand reduction to two-loop
  five-point scattering amplitudes in QCD}},
  \href{https://doi.org/10.22323/1.303.0006}{\emph{PoS} {\bfseries LL2018}
  (2018) 006} [\href{https://arxiv.org/abs/1807.09709}{{\ttfamily
  1807.09709}}].

\bibitem{Abreu:2018jgq}
S.~Abreu, F.~Febres~Cordero, H.~Ita, B.~Page and V.~Sotnikov, \emph{{Planar
  Two-Loop Five-Parton Amplitudes from Numerical Unitarity}},
  \href{https://doi.org/10.1007/JHEP11(2018)116}{\emph{JHEP} {\bfseries 11}
  (2018) 116} [\href{https://arxiv.org/abs/1809.09067}{{\ttfamily
  1809.09067}}].

\bibitem{Gehrmann:2015bfy}
T.~Gehrmann, J.~Henn and N.~Lo~Presti, \emph{{Analytic form of the two-loop
  planar five-gluon all-plus-helicity amplitude in QCD}},
  \href{https://doi.org/10.1103/PhysRevLett.116.062001}{\emph{Phys. Rev. Lett.}
  {\bfseries 116} (2016) 062001}
  [\href{https://arxiv.org/abs/1511.05409}{{\ttfamily 1511.05409}}].

\bibitem{Badger:2018enw}
S.~Badger, C.~Br\o{}nnum-Hansen, H.B.~Hartanto and T.~Peraro, \emph{{Analytic
  helicity amplitudes for two-loop five-gluon scattering: the single-minus
  case}}, \href{https://doi.org/10.1007/JHEP01(2019)186}{\emph{JHEP} {\bfseries
  01} (2019) 186} [\href{https://arxiv.org/abs/1811.11699}{{\ttfamily
  1811.11699}}].

\bibitem{Abreu:2018zmy}
S.~Abreu, J.~Dormans, F.~Febres~Cordero, H.~Ita and B.~Page, \emph{{Analytic
  Form of Planar Two-Loop Five-Gluon Scattering Amplitudes in QCD}},
  \href{https://doi.org/10.1103/PhysRevLett.122.082002}{\emph{Phys. Rev. Lett.}
  {\bfseries 122} (2019) 082002}
  [\href{https://arxiv.org/abs/1812.04586}{{\ttfamily 1812.04586}}].

\bibitem{Abreu:2019odu}
S.~Abreu, J.~Dormans, F.~Febres~Cordero, H.~Ita, B.~Page and V.~Sotnikov,
  \emph{{Analytic Form of the Planar Two-Loop Five-Parton Scattering Amplitudes
  in QCD}}, \href{https://doi.org/10.1007/JHEP05(2019)084}{\emph{JHEP}
  {\bfseries 05} (2019) 084}
  [\href{https://arxiv.org/abs/1904.00945}{{\ttfamily 1904.00945}}].

\bibitem{DeLaurentis:2020qle}
G.~De~Laurentis and D.~Ma\^\i{}tre, \emph{{Two-Loop Five-Parton Leading-Colour
  Finite Remainders in the Spinor-Helicity Formalism}},
  \href{https://doi.org/10.1007/JHEP02(2021)016}{\emph{JHEP} {\bfseries 02}
  (2021) 016} [\href{https://arxiv.org/abs/2010.14525}{{\ttfamily
  2010.14525}}].

\bibitem{Abreu:2018aqd}
S.~Abreu, L.J.~Dixon, E.~Herrmann, B.~Page and M.~Zeng, \emph{{The two-loop
  five-point amplitude in $\mathcal{N} =4$ super-Yang-Mills theory}},
  \href{https://doi.org/10.1103/PhysRevLett.122.121603}{\emph{Phys. Rev. Lett.}
  {\bfseries 122} (2019) 121603}
  [\href{https://arxiv.org/abs/1812.08941}{{\ttfamily 1812.08941}}].

\bibitem{Chicherin:2018yne}
D.~Chicherin, T.~Gehrmann, J.M.~Henn, P.~Wasser, Y.~Zhang and S.~Zoia,
  \emph{{Analytic result for a two-loop five-particle amplitude}},
  \href{https://doi.org/10.1103/PhysRevLett.122.121602}{\emph{Phys. Rev. Lett.}
  {\bfseries 122} (2019) 121602}
  [\href{https://arxiv.org/abs/1812.11057}{{\ttfamily 1812.11057}}].

\bibitem{Caron-Huot:2020vlo}
S.~Caron-Huot, D.~Chicherin, J.~Henn, Y.~Zhang and S.~Zoia, \emph{{Multi-Regge
  Limit of the Two-Loop Five-Point Amplitudes in $\mathcal{N} = 4$ Super
  Yang-Mills and $\mathcal{N} = 8$ Supergravity}},
  \href{https://doi.org/10.1007/JHEP10(2020)188}{\emph{JHEP} {\bfseries 10}
  (2020) 188} [\href{https://arxiv.org/abs/2003.03120}{{\ttfamily
  2003.03120}}].

\bibitem{Badger:2019djh}
S.~Badger, D.~Chicherin, T.~Gehrmann, G.~Heinrich, J.~Henn, T.~Peraro et~al.,
  \emph{{Analytic form of the full two-loop five-gluon all-plus helicity
  amplitude}},
  \href{https://doi.org/10.1103/PhysRevLett.123.071601}{\emph{Phys. Rev. Lett.}
  {\bfseries 123} (2019) 071601}
  [\href{https://arxiv.org/abs/1905.03733}{{\ttfamily 1905.03733}}].

\bibitem{Berger:2009ep}
C.F.~Berger, Z.~Bern, L.J.~Dixon, F.~Febres~Cordero, D.~Forde, T.~Gleisberg
  et~al., \emph{{Next-to-Leading Order QCD Predictions for W+3-Jet
  Distributions at Hadron Colliders}},
  \href{https://doi.org/10.1103/PhysRevD.80.074036}{\emph{Phys. Rev. D}
  {\bfseries 80} (2009) 074036}
  [\href{https://arxiv.org/abs/0907.1984}{{\ttfamily 0907.1984}}].

\bibitem{Ita:2011ar}
H.~Ita and K.~Ozeren, \emph{{Colour Decompositions of Multi-quark One-loop QCD
  Amplitudes}}, \href{https://doi.org/10.1007/JHEP02(2012)118}{\emph{JHEP}
  {\bfseries 02} (2012) 118} [\href{https://arxiv.org/abs/1111.4193}{{\ttfamily
  1111.4193}}].

\bibitem{Abreu:2020cwb}
S.~Abreu, B.~Page, E.~Pascual and V.~Sotnikov, \emph{{Leading-Color Two-Loop
  QCD Corrections for Three-Photon Production at Hadron Colliders}},
  \href{https://doi.org/10.1007/JHEP01(2021)078}{\emph{JHEP} {\bfseries 21}
  (2020) 078} [\href{https://arxiv.org/abs/2010.15834}{{\ttfamily
  2010.15834}}].

\bibitem{Chawdhry:2020for}
H.A.~Chawdhry, M.~Czakon, A.~Mitov and R.~Poncelet, \emph{{Two-loop
  leading-color helicity amplitudes for three-photon production at the LHC}},
  \href{https://arxiv.org/abs/2012.13553}{{\ttfamily 2012.13553}}.

\bibitem{Chawdhry:2019bji}
H.A.~Chawdhry, M.L.~Czakon, A.~Mitov and R.~Poncelet, \emph{{NNLO QCD
  corrections to three-photon production at the LHC}},
  \href{https://doi.org/10.1007/JHEP02(2020)057}{\emph{JHEP} {\bfseries 02}
  (2020) 057} [\href{https://arxiv.org/abs/1911.00479}{{\ttfamily
  1911.00479}}].

\bibitem{Kallweit:2020gcp}
S.~Kallweit, V.~Sotnikov and M.~Wiesemann, \emph{{Triphoton production at
  hadron colliders in NNLO QCD}},
  \href{https://doi.org/10.1016/j.physletb.2020.136013}{\emph{Phys. Lett. B}
  {\bfseries 812} (2021) 136013}
  [\href{https://arxiv.org/abs/2010.04681}{{\ttfamily 2010.04681}}].

\bibitem{Agarwal:2021grm}
B.~Agarwal, F.~Buccioni, A.~von Manteuffel and L.~Tancredi, \emph{{Two-loop
  leading colour QCD corrections to $q \bar{q} \to \gamma \gamma g$ and $q g
  \to \gamma \gamma q$}},  \href{https://arxiv.org/abs/2102.01820}{{\ttfamily
  2102.01820}}.

\bibitem{Badger:2021nhg}
S.~Badger, H.B.~Hartanto and S.~Zoia, \emph{{Two-loop QCD corrections to
  $Wb\bar{b}$ production at hadron colliders}},
  \href{https://arxiv.org/abs/2102.02516}{{\ttfamily 2102.02516}}.

\bibitem{Papadopoulos:2015jft}
C.G.~Papadopoulos, D.~Tommasini and C.~Wever, \emph{{The Pentabox Master
  Integrals with the Simplified Differential Equations approach}},
  \href{https://doi.org/10.1007/JHEP04(2016)078}{\emph{JHEP} {\bfseries 04}
  (2016) 078} [\href{https://arxiv.org/abs/1511.09404}{{\ttfamily
  1511.09404}}].

\bibitem{Chicherin:2018old}
D.~Chicherin, T.~Gehrmann, J.~Henn, P.~Wasser, Y.~Zhang and S.~Zoia, \emph{{All
  Master Integrals for Three-Jet Production at Next-to-Next-to-Leading Order}},
  \href{https://doi.org/10.1103/PhysRevLett.123.041603}{\emph{Phys. Rev. Lett.}
  {\bfseries 123} (2019) 041603}
  [\href{https://arxiv.org/abs/1812.11160}{{\ttfamily 1812.11160}}].

\bibitem{Gehrmann:2018yef}
T.~Gehrmann, J.~Henn and N.~Lo~Presti, \emph{{Pentagon functions for massless
  planar scattering amplitudes}},
  \href{https://doi.org/10.1007/JHEP10(2018)103}{\emph{JHEP} {\bfseries 10}
  (2018) 103} [\href{https://arxiv.org/abs/1807.09812}{{\ttfamily
  1807.09812}}].

\bibitem{Chicherin:2020oor}
D.~Chicherin and V.~Sotnikov, \emph{{Pentagon Functions for Scattering of Five
  Massless Particles}},
  \href{https://doi.org/10.1007/JHEP12(2020)167}{\emph{JHEP} {\bfseries 12}
  (2020) 167} [\href{https://arxiv.org/abs/2009.07803}{{\ttfamily
  2009.07803}}].

\bibitem{vonManteuffel:2014ixa}
A.~von Manteuffel and R.M.~Schabinger, \emph{{A novel approach to integration
  by parts reduction}},
  \href{https://doi.org/10.1016/j.physletb.2015.03.029}{\emph{Phys. Lett. B}
  {\bfseries 744} (2015) 101}
  [\href{https://arxiv.org/abs/1406.4513}{{\ttfamily 1406.4513}}].

\bibitem{Peraro:2016wsq}
T.~Peraro, \emph{{Scattering amplitudes over finite fields and multivariate
  functional reconstruction}},
  \href{https://doi.org/10.1007/JHEP12(2016)030}{\emph{JHEP} {\bfseries 12}
  (2016) 030} [\href{https://arxiv.org/abs/1608.01902}{{\ttfamily
  1608.01902}}].

\bibitem{Abreu:2020xvt}
S.~Abreu, J.~Dormans, F.~Febres~Cordero, H.~Ita, M.~Kraus, B.~Page et~al.,
  \emph{{Caravel: A C++ Framework for the Computation of Multi-Loop Amplitudes
  with Numerical Unitarity}},
  \href{https://doi.org/https://doi.org/10.1016/j.cpc.2021.108069}{\emph{Computer
  Physics Communications} {\bfseries 267} (2021) 108069}
  [\href{https://arxiv.org/abs/2009.11957}{{\ttfamily 2009.11957}}].

\bibitem{FivePointAmplitudes}
\url{https://gitlab.com/five-point-amplitudes/FivePointAmplitudes-cpp.git}.

\bibitem{Weinzierl:2011uz}
S.~Weinzierl, \emph{{Does one need the O(epsilon)- and
  O(epsilon\textasciicircum{}2)-terms of one-loop amplitudes in an NNLO
  calculation ?}},
  \href{https://doi.org/10.1103/PhysRevD.84.074007}{\emph{Phys. Rev. D}
  {\bfseries 84} (2011) 074007}
  [\href{https://arxiv.org/abs/1107.5131}{{\ttfamily 1107.5131}}].

\bibitem{Catani:1998bh}
S.~Catani, \emph{{The Singular behavior of QCD amplitudes at two loop order}},
  \href{https://doi.org/10.1016/S0370-2693(98)00332-3}{\emph{Phys. Lett. B}
  {\bfseries 427} (1998) 161}
  [\href{https://arxiv.org/abs/hep-ph/9802439}{{\ttfamily hep-ph/9802439}}].

\bibitem{Becher:2009cu}
T.~Becher and M.~Neubert, \emph{{Infrared singularities of scattering
  amplitudes in perturbative QCD}},
  \href{https://doi.org/10.1103/PhysRevLett.102.162001}{\emph{Phys. Rev. Lett.}
  {\bfseries 102} (2009) 162001}
  [\href{https://arxiv.org/abs/0901.0722}{{\ttfamily 0901.0722}}].

\bibitem{Gardi:2009qi}
E.~Gardi and L.~Magnea, \emph{{Factorization constraints for soft anomalous
  dimensions in QCD scattering amplitudes}},
  \href{https://doi.org/10.1088/1126-6708/2009/03/079}{\emph{JHEP} {\bfseries
  03} (2009) 079} [\href{https://arxiv.org/abs/0901.1091}{{\ttfamily
  0901.1091}}].

\bibitem{Broggio:2015dga}
A.~Broggio, C.~Gnendiger, A.~Signer, D.~St\"ockinger and A.~Visconti,
  \emph{{SCET approach to regularization-scheme dependence of QCD amplitudes}},
  \href{https://doi.org/10.1007/JHEP01(2016)078}{\emph{JHEP} {\bfseries 01}
  (2016) 078} [\href{https://arxiv.org/abs/1506.05301}{{\ttfamily
  1506.05301}}].

\bibitem{Ita:2015tya}
H.~Ita, \emph{{Two-loop Integrand Decomposition into Master Integrals and
  Surface Terms}},
  \href{https://doi.org/10.1103/PhysRevD.94.116015}{\emph{Phys. Rev. D}
  {\bfseries 94} (2016) 116015}
  [\href{https://arxiv.org/abs/1510.05626}{{\ttfamily 1510.05626}}].

\bibitem{Abreu:2017xsl}
S.~Abreu, F.~Febres~Cordero, H.~Ita, M.~Jaquier, B.~Page and M.~Zeng,
  \emph{{Two-Loop Four-Gluon Amplitudes from Numerical Unitarity}},
  \href{https://doi.org/10.1103/PhysRevLett.119.142001}{\emph{Phys. Rev. Lett.}
  {\bfseries 119} (2017) 142001}
  [\href{https://arxiv.org/abs/1703.05273}{{\ttfamily 1703.05273}}].

\bibitem{ArkaniHamed:2010gh}
N.~Arkani-Hamed, J.L.~Bourjaily, F.~Cachazo and J.~Trnka, \emph{{Local
  Integrals for Planar Scattering Amplitudes}},
  \href{https://doi.org/10.1007/JHEP06(2012)125}{\emph{JHEP} {\bfseries 06}
  (2012) 125} [\href{https://arxiv.org/abs/1012.6032}{{\ttfamily 1012.6032}}].

\bibitem{Ruijl:2017dtg}
B.~Ruijl, T.~Ueda and J.~Vermaseren, \emph{{FORM version 4.2}},
  \href{https://arxiv.org/abs/1707.06453}{{\ttfamily 1707.06453}}.

\bibitem{Kuipers:2013pba}
J.~Kuipers, T.~Ueda and J.~Vermaseren, \emph{{Code Optimization in FORM}},
  \href{https://doi.org/10.1016/j.cpc.2014.08.008}{\emph{Comput. Phys. Commun.}
  {\bfseries 189} (2015) 1} [\href{https://arxiv.org/abs/1310.7007}{{\ttfamily
  1310.7007}}].

\bibitem{Chicherin:2017dob}
D.~Chicherin, J.~Henn and V.~Mitev, \emph{{Bootstrapping pentagon functions}},
  \href{https://doi.org/10.1007/JHEP05(2018)164}{\emph{JHEP} {\bfseries 05}
  (2018) 164} [\href{https://arxiv.org/abs/1712.09610}{{\ttfamily
  1712.09610}}].

\bibitem{Bothmann:2019yzt}
{\scshape Sherpa} collaboration, \emph{{Event Generation with Sherpa 2.2}},
  \href{https://doi.org/10.21468/SciPostPhys.7.3.034}{\emph{SciPost Phys.}
  {\bfseries 7} (2019) 034} [\href{https://arxiv.org/abs/1905.09127}{{\ttfamily
  1905.09127}}].

\bibitem{Aad:2014rma}
{\scshape ATLAS} collaboration, \emph{{Measurement of three-jet production
  cross-sections in $pp$ collisions at 7 TeV centre-of-mass energy using the
  ATLAS detector}},
  \href{https://doi.org/10.1140/epjc/s10052-015-3363-3}{\emph{Eur. Phys. J. C}
  {\bfseries 75} (2015) 228} [\href{https://arxiv.org/abs/1411.1855}{{\ttfamily
  1411.1855}}].

\bibitem{Cacciari:2011ma}
M.~Cacciari, G.P.~Salam and G.~Soyez, \emph{{FastJet User Manual}},
  \href{https://doi.org/10.1140/epjc/s10052-012-1896-2}{\emph{Eur. Phys. J. C}
  {\bfseries 72} (2012) 1896}
  [\href{https://arxiv.org/abs/1111.6097}{{\ttfamily 1111.6097}}].

\bibitem{Badger:correspondence}
S.~Badger, C.~Br\o{}nnum-Hansen, H.B.~Hartanto, R.~Moodie, T.~Peraro and
  S.~Zoia. private correspondence, 2021.

\bibitem{Berger:2008sj}
C.F.~Berger, Z.~Bern, L.J.~Dixon, F.~Febres~Cordero, D.~Forde, H.~Ita et~al.,
  \emph{{An Automated Implementation of On-Shell Methods for One-Loop
  Amplitudes}}, \href{https://doi.org/10.1103/PhysRevD.78.036003}{\emph{Phys.
  Rev.} {\bfseries D78} (2008) 036003}
  [\href{https://arxiv.org/abs/0803.4180}{{\ttfamily 0803.4180}}].

\end{thebibliography}\endgroup

\end{document}